# Fine-Tuning Hydrophilic-Hydrophobic Balance in Stimuli-Responsive PEG-PNIPAM Micelles for Controlled Drug Delivery

G. Jamirad[1], M. Seif[1], A. Montazeri[1, 2,*]

[1]Computational Nanomaterials Lab (CNL), Faculty of Materials Science and Engineering, , K. N. Toosi University of Technology, Tehran, Iran
[2]School of Quantum Physics and Matter, Institute for Research in Fundamental Sciences, IPM, P.O. 19395-5531, Tehran, Iran

## Abstract

Temperature-responsive polymeric micelles have great potential in drug delivery, enhancing therapeutic efficacy while minimizing systemic toxicity. This study investigates the self-assembly of poly(ethylene glycol)-poly(N-isopropylacrylamide) (PEG-PNIPAM) block copolymers as nanocarriers for Doxorubicin (DOX) using coarse-grained molecular dynamics (CG-MD) simulations. We systematically examined the effects of polymer chain length and hydrophilic/hydrophobic balance on micelle formation, stability, and drug encapsulation efficiency. PEG-PNIPAM copolymers self-assembled above the lower critical solution temperature (LCST), forming stable micelles with varying morphologies. Our findings revealed that PEG enhances micelle hydration and stability, while PNIPAM improves drug retention but may hinder controlled release. Increasing PNIPAM or PEG length significantly increased micelle size, with DOX preferentially localizing in the PEG-rich outer shell, enhancing solubility and preventing aggregation. The DOX diffusion coefficient ranged from $10^{-8}$ to $10^{-9}$ cm²·s⁻¹, reflecting drug release dynamics. Structural and thermodynamic analyses confirmed spontaneous DOX encapsulation, with PEG-PNIPAM micelles providing a favorable environment for drug loading. Mean square displacement (MSD), radial distribution function (RDF), and solvent-accessible surface area (SASA) analyses highlighted PEG's role in controlled drug release and PNIPAM's role in drug binding. These molecular-level insights underscore the potential of fine-tuning PEG-PNIPAM compositions for enhanced drug delivery, paving the way for precision cancer therapies.

## Keywords:

* Corresponding author:

mailto:a_montazeri@kntu.ac.ir (A. Montazeri).

# 1. Introduction

Cancer is the leading cause of death worldwide and a major global burden, making its prevention and treatment a top priority. Doxorubicin (DOX), a key chemotherapy drug from the anthracycline family ($C_{27}H_{29}NO_{11}$), is effective against various cancers [1]. However, DOX has a short biological lifetime and elicits adverse side effects, including cardiotoxicity, nausea, and vomiting, induced by its nonspecific biodistribution [2]. A promising approach to improve the solubility and effectiveness of amphiphilic drugs like DOX is incorporating them into nanocarriers. Self-assembled polymeric micelles are increasingly recognized as promising drug delivery systems (DDSs), particularly in the field of cancer therapy. These systems are utilized as carriers for various therapeutic agents, including cancer drugs, to address the significant challenges associated with traditional anticancer treatments such as poor bioavailability, low solubility, uncontrolled drug release, untargeted delivery, and severe side effects [3]. Micelles formed in aqueous media by the hydrophobic cores of block copolymers serve as ideal reservoirs for encapsulating DOX for controlled release, while their hydrophilic outer shells provide stability [4]. The enhanced permeability and retention (EPR) effect and prolonged circulation time achieved by micellar delivery systems alleviate DOX side effects and improve therapeutic efficiency.

Stimuli-responsive drug delivery systems, which release drugs in response to external or internal stimuli (such as temperature, pH, and light), have emerged as an effective strategy to improve targeted drug delivery and to facilitate drug release. In this context, biocompatible block copolymer assemblies that respond to temperature and pH changes have shown promise since tumors typically exhibit higher temperatures and acidity compared to normal tissues. Among these, poly(N-isopropyl acrylamide) (PNIPAM) is a commonly used thermosensitive polymer that undergoes a phase transition at its lower critical solution temperature (LCST) (~32°C) [5]. As a result, PNIPAM has gained significant recognition and is widely implemented in various applications, including drug delivery [6], tissue engineering [7], biosensing [8], and shape memory technologies [9]. Approaching the LCST leads to changes in several properties, such as a reversible volume phase transition, hydrophilicity, transparency, and electrostatic permittivity [10]. Below this temperature, PNIPAM remains hydrated and swollen, while above it, the polymer undergoes a coil-to-globule transition to adopt a more compact conformation. This phase transition is crucial for drug delivery applications as it allows for the controlled release of encapsulated drugs [11]. PNIPAM is neither completely hydrophobic nor hydrophilic; both its amide and isopropyl groups retain their respective hydrophilic and hydrophobic characteristics above and below the LCST [12]. This dual nature allows PNIPAM to load both hydrophilic and hydrophobic compounds, making it versatile for drug delivery applications, such as with DOX. A DOX–PNIPAM-co-acrylic acid graft copolymer conjugate has been developed for targeted anticancer drug delivery to breast cancer cells [13]. In vitro studies show that PNIPAM-based structures are effective for sustained drug release, especially in DOX-loaded formulations [14]. These structures enable efficient drug release in aqueous environments, are biocompatible, enhance cell toxicity against melanoma cells, and improve tumor-suppressing ability when exposed to NIR light [15].

The LCST of PNIPAM can be adjusted through copolymerization with other hydrophilic or hydrophobic monomers for better targeting and delivery [16]. These amphiphilic copolymers can self-assemble into nanosized polymeric micelles. Different micelle morphologies, including rods, spheres, and vesicles, are possible depending on the copolymer structure and external conditions like pH, solvent, temperature, and polymer concentration [17]. Two common types of PNIPAM-based micelles are reported: core-shell micelles with a hydrophobic inner core and a hydrated PNIPAM shell, and double hydrophilic block copolymers (DHBCs) where PNIPAM acts as the core above the LCST and is coupled with a hydrophilic block. DHBCs can dissolve in aqueous media below the LCST and self-assemble into colloidal micelles above the LCST due to the hydrophilic-to-hydrophobic transformation of PNIPAM at higher temperatures [18]. The desired hydrophilicity can be achieved by incorporating Poly(ethylene glycol) (PEG) moieties into the hydrophobic polymer segments. PEG is a linear, nonionic, hydrophilic, biodegradable, and biocompatible polymer widely studied for its role in active pharmaceutical ingredient (API) release properties [19]. PEG plays a crucial role in the stabilization of micelles due to its highly hydrated features, providing stealth properties that help them evade the immune system and extend circulation time in the bloodstream [20]. Micelles formed by PEG-PNIPAM copolymers have been explored for encapsulating lipophilic drugs, with temperature-triggered disassembly enabling controlled drug release. These micelles possess a hydrophilic PEG shell and a thermosensitive PNIPAM core. At temperatures above the LCST, the PNIPAM core can encapsulate lipophilic drugs, and upon cooling below the LCST, a burst drug release can be achieved through the disassembly of the micelle structure. This temperature-controlled release mechanism makes these micelles suitable for smart drug delivery applications. The block copolymer composition, including block length and the hydrophilic/hydrophobic ratio, is crucial for optimizing polymeric micelles in drug delivery. Previous studies have demonstrated that micelle stability and drug delivery depend on these factors, with polymer chain length influencing shape, size, and accessible surface area of the micelles [21, 22]. Manipulating chain length affects amphiphilic copolymer self-assembly [22], with surface block length impacting biological behavior [23]. Additionally, PEG chain length can influence micelle stability, stealth properties, biodistribution, and cellular uptake [24]. Research conducted by Cremer's group has also explored how salt, chain length, and end-group polarity affect PNIPAM's LCST and solubility [25,26].

Traditional experimental techniques for analyzing DDSs, such as electron microscopy and mass spectrometry, offer structural insights but are limited in their ability to capture dynamic processes of formation and evolution [27]. Computational methods, especially molecular dynamics (MD) simulations, are increasingly utilized to understand molecular-level interactions within DDSs [28]. MD simulations provide valuable insights into micellar self-assembly and carrier morphology [29], drug encapsulation [30], aggregation and toxicity mechanisms [31], and interactions between drugs and carriers, information that may not always be accessible through experimental means. These simulation studies are crucial in DDS development, offering a controlled, virtual environment to explore system behavior and performance, ultimately aiding in the rational design of DDSs. However, simulating large systems, like polymeric micelles, with all-atom (AA) models is computationally demanding, especially for long simulations (~1 μs) and large system sizes (~100 nm). To reduce complexity while preserving key chemical properties, coarse-grained (CG) models are often employed. CG molecular dynamics (CG-MD) simulations simplify molecule representation, allowing for longer simulations and larger systems. This approach has been successfully applied to study self-assembly and vesicle formation processes, offering valuable insights into the behavior of polymeric micelles. For example, Duran et al. [32], applied CG-MD to investigate spontaneous micellar self-assembly, while Mobasheri et al. [33] analyzed nanomicellar solution structures in detail, capturing key features of the formation process.

The properties of polymeric micelles (e.g., drug loading capacity, stability, and release rate) are directly influenced by their structures. Optimizing micelle structures is crucial for enhancing DDSs, yet studies on the structure-property relationship of stimuli-responsive polymeric micelles are limited due to their nanoscale complexity. Developing non-toxic, targeted, and controlled-release therapeutic strategies for DOX is essential, as smart DDSs offer clear advantages over conventional methods. PNIPAM polymers, which undergo conformational changes from flexible coils to globular states, play a key role in drug absorption and release. However, challenges remain, including lower-than-expected delivery efficiency due to interactions with biological environments and the complexity of designing block copolymers. Critical quality attributes, such as drug loading, encapsulation efficiency, and stability, are significantly affected by material properties, including the hydrophilic-to-hydrophobic ratio and polymer-drug interactions. Using MD simulation techniques, this study systematically examines the balance between hydrophobic NIPAM (above LCST) and hydrophilic PEG in self-assembly and DOX entrapment, a topic not yet fully explored. To identify the optimal hydrophobic/hydrophilic balance, we prepare a series of copolymers with varying monomer ratios and use coarse-graining techniques to simulate the self-assembly process and DOX encapsulation within PEG-PNIPAM micelles. CG-MD simulations reveal the polymer aggregate structure and characterize it using various analysis methods. Our findings provide valuable insights into the stability and physiological properties of PEG-NIPAM micelles, with potential applications in targeted cancer therapies.

## 2. Methodology and Simulation Details

This study investigates the aggregation of DOX, the self-assembly of PEG-PNIPAM-based micelles in an aqueous medium, and the encapsulation process of DOX within micelles. Key areas of focus include intermolecular interactions during drug encapsulation, structural properties of PEG-PNIPAM nano-micelles, drug diffusion behavior, and the stability of the drug-nanocarrier mixed micellar system. The following sections outline the modeling approaches and simulation methodologies employed.

### 2.1. Atomistic Modeling and Simulation of Copolymers

Figure 1a illustrates the chemical structure of a block copolymer composed of PEG-PNIPAM monomers. A bottom-up modeling strategy was employed to determine the CG structural parameters of the PEG-PNIPAM copolymer. AA simulations were used to generate the necessary parameters to characterize the bonded interactions within the CG models. The copolymer chain length was chosen based on factors such as DOX delivery efficiency [34], medical applications [35], solubility, biocompatibility [36,37], temperature responsiveness [34], toxicity [38], and computational cost. To assess the influence of PNIPAM and PEG block lengths on micellar properties, copolymer chains with 10 and 20 PEG monomers, alongside 20, 25, 30, and 40 PNIPAM monomers, were created. The systems are labeled as PEG*m*NIP*n*, where *m* and *n* represent the number of PEG and PNIPAM monomers, respectively. For instance, PEG10NIP20 has 10 PEG and 20 PNIPAM monomers. Most PNIPAM synthesis methods produce atactic chains [39], and simulations generally adopt this stereochemistry [40]. Thus, the AA polymer chain was constructed as an atactic structure using CHARMM-GUI (http://www.charmm-gui.org). The OPLS-AA force field was applied to parameterize and model AA PEG and PNIPAM monomers, in conjunction with the SPC/E water model [41,42]. All AA and CG simulations were conducted using the GROMACS 2022 software package [43]. The simulation began by constructing a single polymer chain within a box and adding water molecules to simulate an aqueous environment. As shown in Fig. 2, the system underwent minimization using the steepest descent method, followed by NVT equilibration (200 ps) and NPT ensemble (200 ps) to relax the system. Following this, 1 ns production runs were performed using the Velocity-Rescaling thermostat [44] and Parrinello-Rahman barostat [45] at T=300 K and P=1 bar, with a 2 fs time-step and a 1.2 nm cutoff through the leapfrog integrator. The output of these simulations served as the basis for the CG simulations.

a)                                              b)

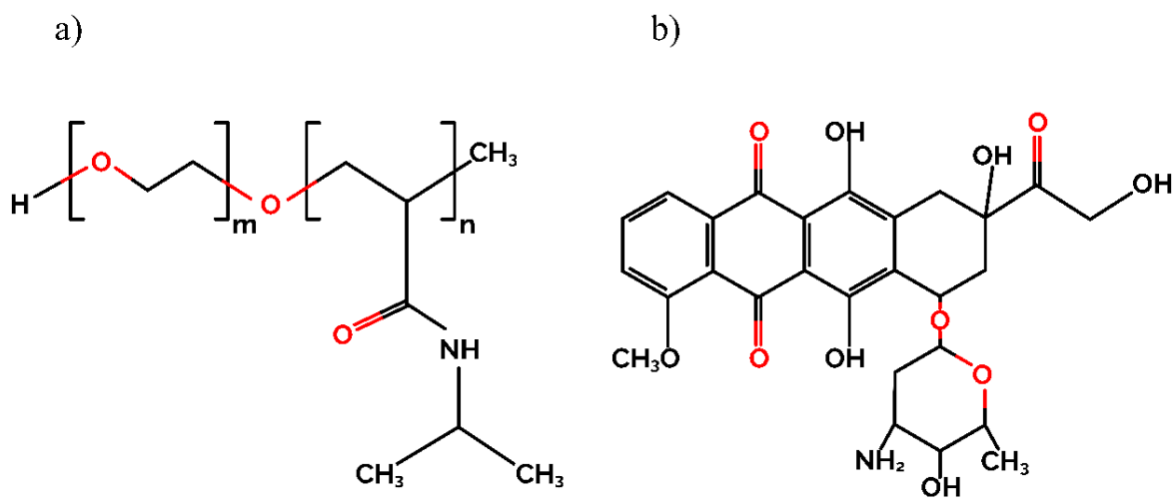

Fig. 1. Chemical Structure of (a) PEG-PNIPAM and (b) Doxorubicin.

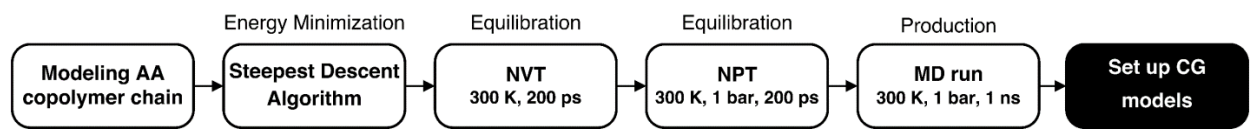

Fig. 2. Overview of the AA simulation protocol for energy minimization, equilibration of systems, and derivation of CG parameters.

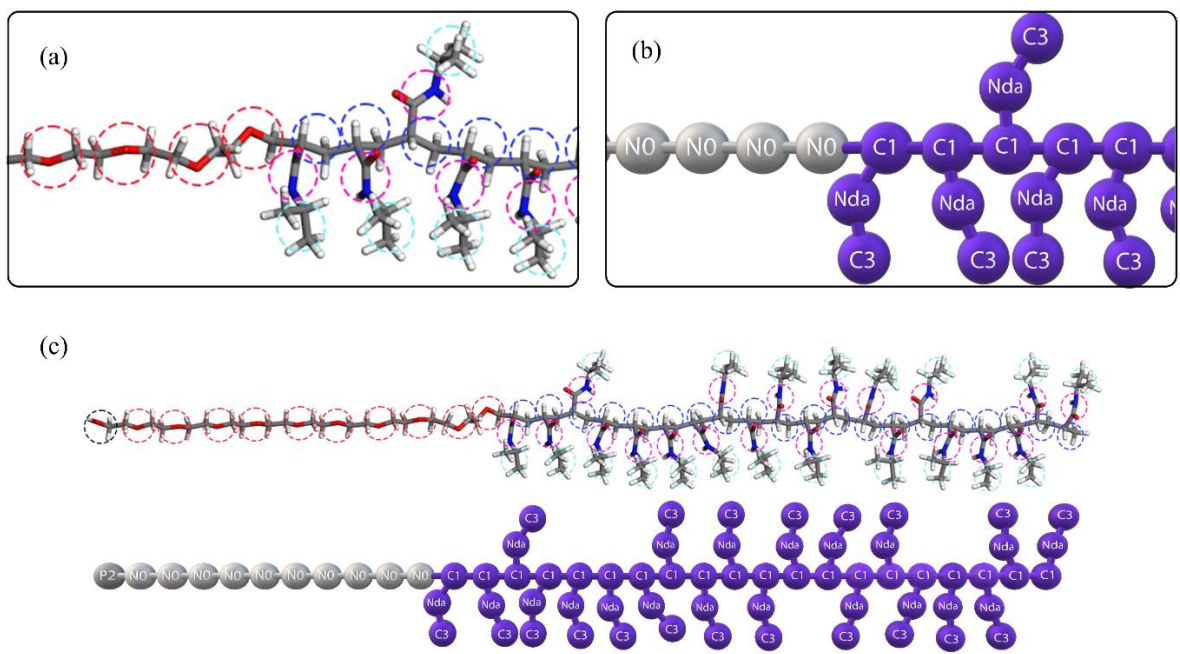

Fig. 3. Coarse-grained bead representation of the PEG-PNIPAM copolymer chain with PEG10NIP20 as a representative sample: (a) AA structure and (b) CG structure, and (c) coarse-grained PEG10NIP20 chain.

## 2.2. Coarse-grained Modeling and Simulation of Copolymers

CG models were initially derived from the corresponding all-atom configurations at the relevant temperatures. A widely used approach in CG-MD simulations is the MARTINI method, developed and extended by Marrink et al. [46]. In this method, PEG-PNIPAM blocks and DOX were separately mapped using the standard MARTINI coarse-graining strategy (see Figs. 3 & 4). The characteristics of the PEG-PNIPAM copolymer are preserved in the CG representation, as established by prior analysis and published studies. Specifically, the "N0" bead is assigned to the ethylene glycol monomer, while the "P2" bead represents the OH-terminal group [32,47,48]. The PNIPAM monomer is represented by "C1," "Nda," and "C3" beads, which correspond to the hydrophilic and hydrophobic components, respectively [41]. The nonbonded interactions for these mapped beads in the MARTINI force field are represented by Lennard-Jones (LJ) potential functions. Periodic boundary conditions (PBCs) were assigned in all dimensions.

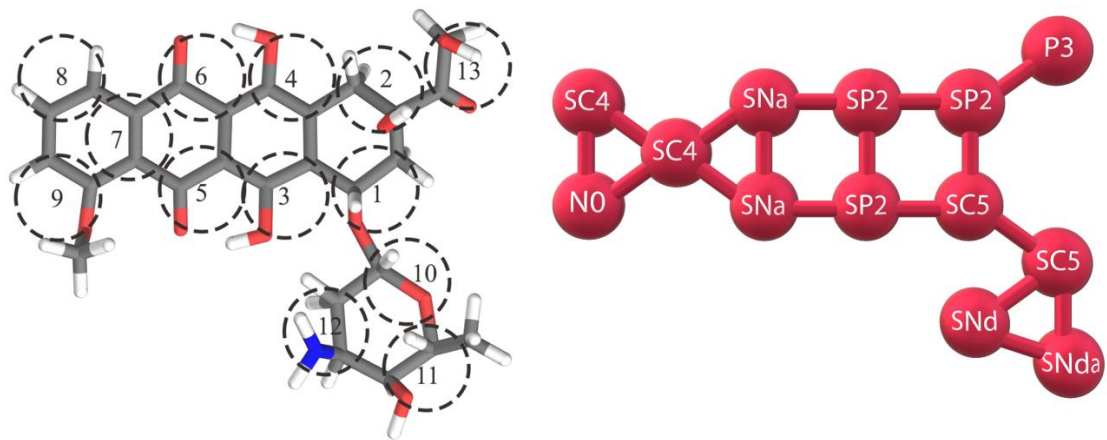

Fig. 4. Coarse-grained bead representation of the Doxorubicin molecules.

The experimental method for PEG-PNIPAM micelles, which forms the basis of these simulations, involves heating a polymer solution above the LCST to form micelles in aqueous media without the use of organic solvents [49]. The temperature was set at 310 K (above LCST) to simulate the conditions of an aqueous medium in the human body. The self-assembly of blank micelles was simulated using 100 CG chains at a concentration of 10 wt.% [4,37,38]. Initially, the polymer chains were randomly distributed in cubic simulation boxes, with their sizes as shown in Table 1. To mitigate the potential adverse effects of CG water, 10% of the molecules were replaced with antifreeze water (BW) molecules [46].

Table 1. Summary of simulation parameters for blank micelles, including degree of polymerization (DP), molecular weights of copolymers, number of water molecules, box size, and simulation time

| Sample | Degree of Polymerization | Molecular Weight (g.mol⁻¹) | # Water Molecules | Simulation Box Size (nm³) | Simulation Time (ns) |
|---|---|---|---|---|---|
| NIP25 | 100 | 2827 | 35308 | 17.4×17.4×17.4 | 700 |
| PEG10NIP20 | 100 | 3012 | 37618 | 17.7×17.7×17.7 | 700 |
| PEG10NIP25 | 100 | 3297 | 41178 | 18.2×18.2×18.2 | 700 |
| PEG10NIP30 | 100 | 4282 | 53480 | 19.9×19.9×19.9 | 700 |
| PEG10NIP40 | 100 | 5552 | 69341 | 21.7×21.7×21.7 | 700 |
| PEG20NIP25 | 100 | 3737 | 46672 | 19.1×19.1×19.1 | 700 |

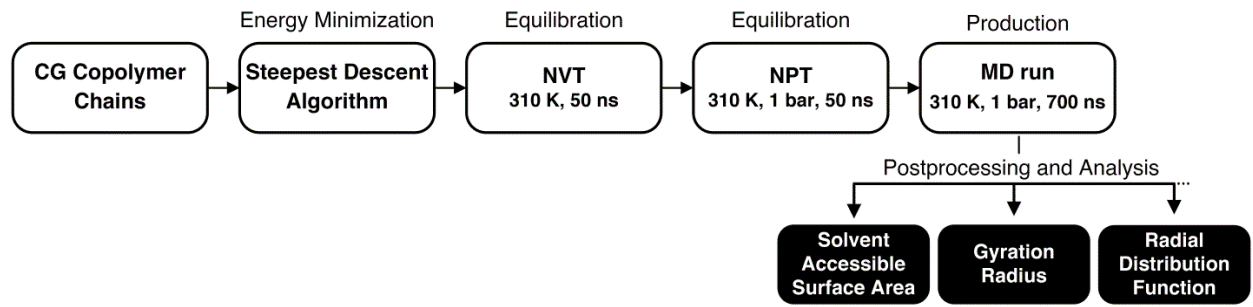

Fig. 5. Overview of the CG simulation protocol for energy minimization, system equilibration, and production runs.

As depicted in Fig. 5, before the production runs and subsequent analysis, the system underwent an equilibration and energy minimization process. A 700 ns production run was then performed with a velocity-rescaling thermostat [44] and Berendsen barostat [50] at T = 310 K and P = 1 bar, with a time step of 20 fs. The leapfrog algorithm was used to integrate Newton's equations of motion. A standard Verlet cutoff scheme, with a cutoff of 1.2 nm, was applied to nonbonded interactions. To evaluate the key properties, post-processing and analysis tools were employed, including final trajectory snapshots, radius of gyration (Rg), solvent accessible surface area (SASA), principal moments of inertia (PMI), density distribution, radial distribution function (RDF), and mean square displacement (MSD). The dynamic data were analyzed using the GROMACS software package, while the simulated snapshots were visualized with the VMD software [51]. Each simulation was repeated three to five times to account for any variability in the results. This approach allowed for the calculation of the standard error, ensuring that any slight differences in the reported values fell within the confidence intervals. Details regarding the modeling and simulation of the drug-loaded micelles are provided in the Supporting Information.

## 3. Results and Discussion

A systematic study was conducted to examine how polymer chain length affects self-assembly, with the goal of establishing an optimal copolymer composition. Computational models were validated, and the self-assembly of PEG-PNIPAM micelles was explored in both blank and drug-loaded systems.

### 3.1. Model Validation and Accuracy Assessments

The accuracy of our computational models was validated using a standard glass transition test, coil-to-globule transition analysis, and physical property monitoring. System temperature, pressure, potential, and density were monitored prior to MD simulations to confirm system equilibration (Figure S.1 in the Supporting Information). After minimizing and equilibrating the CG models, the final polymer density of 100 chains of NIP25 was found to be 0.9623 g.cm$^{-3}$, which aligns well with experimental and MD results (0.90-0.95 g.cm$^{-3}$) [37,52]. Polymers typically transition from a glassy (crystalline) state to a rubbery one as temperature increases [53]. The temperature at which this transition occurs is known as the glass transition temperature (Tg). An increase in polymer chain mobility above Tg can result in abrupt changes in polymer properties. Tg can be determined by fitting a curve to the density versus temperature, identifying the intersection of asymptotes at high and low temperatures. Our goal was to determine Tg for PNIPAM after equilibration by monitoring the specific volume of the polymer models as a function of temperature at constant pressure. Using a time-step of 20 fs, we gradually reduced the CG NIP25 temperature from 500 to 250 K, resulting in an abrupt change in the slope of the curve while applying PBCs in all directions. This method has been successfully applied in previous studies [32,54]. The fitting (Fig. 6a) yielded a Tg value for the CG PNIPAM model of 393.15 K, which is close to the average experimentally measured value of 406.1 K [55,56].

A series of AA simulations using validated OPLS-AA parameters was performed to derive MARTINI CG beads and bonds. Proper evaluations ensured the successful transfer to the CG force field. The key feature of PNIPAM polymer is its phase transition temperature, which is within the range of human body temperature [37]. A PEG10NIP20 chain in water was simulated at six temperatures (280, 290, 300, 310, 320, and 330 K) to examine its coil-to-globule transition behavior, consistent with the procedure shown in Fig. 5. Figure 6b shows that polymer chains are linear at 280 K but become globular at 330 K. The sharp decrease in Rg confirms the model's accurate representation of the chain's thermo-responsive behavior [34]. We compared the calculated Rg and SASA values with those from previous AA simulations and experimental studies. The average Rg of 26 monomers of PNIPAM in the AA models of Moghadam et al. [39], simulated over 300 ns at 310 K, is similar to the Rg of a single NIP25 chain in our study (1.563 ± 0.11 nm). Additionally, our calculated Rg and SASA for a single PEG10NIP20 chain (1.19 ± 0.17 nm and 33.35 ± 2.39 nm²) are in reasonable agreement with the AA simulation results of Rezaeisadat et al. [37], which reported values of 0.99 ± 0.06 nm and 29.0 ± 0.07 nm² for a PEG5NIP20 single chain at 310 K. Furthermore, Rg for a PNIPAM chain with 30 monomers was previously reported as 1.68 ± 0.08 nm based on experimental measurements and AA simulations [57], while our CG model estimated an Rg of 1.86 ± 0.29 nm for the PEG10NIP30 system. The discrepancy is likely due to the addition of 10 monomers of PEG. These results indicate that our CG model provides a reasonable approximation of chain dimensions and solvent exposure, supporting its validity for studying polymeric behavior.

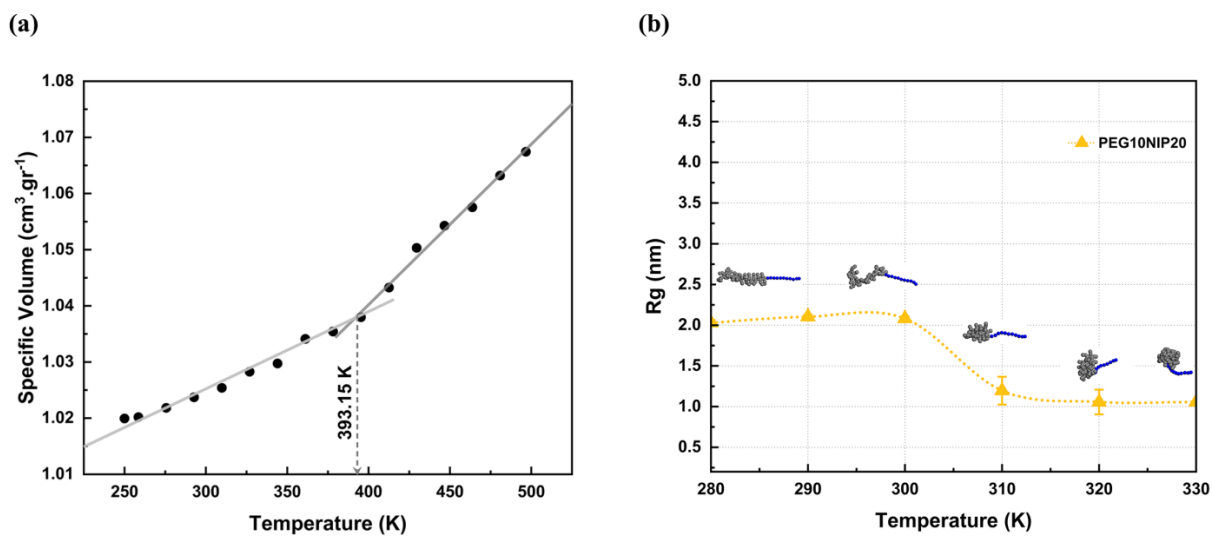

Fig. 6. Model evaluations: (a) Tg of PNIPAM using specific volume vs. temperature; (b) Rg for single polymer chain (PEG10NIP20) at different temperatures (280, 290, 300, 310, 320, and 330 K).

### 3.2. Blank PEG-PNIPAM Self-Assembly at Body Temperature

During 700 ns simulations, we monitored the formation, growth, and phase transition of PEG-PNIPAM systems through trajectory snapshots to investigate blank micelles creation. Key properties, such as SASA, Rg, PMI, RDF, and density distribution, were used to study the self-assembly dynamics, structural characteristics, and conformational changes of the micelles. These analyses provide insights into micelle structure, environmental interactions, and the influence of PEG and PNIPAM composition.

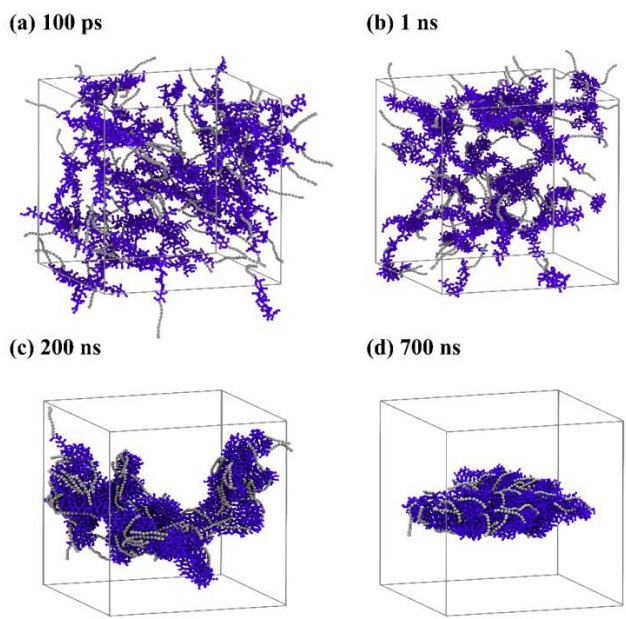

Fig. 7. Simulation snapshots of blank micelle self-assembly in aqueous solvent over time: (a) 100 ps, (b) 1 ns, (c) 200 ns, and (d) 700 ns.

### 3.2.1. Dynamic Assembly Snapshots

Direct observation of spontaneous micelle formation is challenging, but MD simulations allow us to visualize this process. The self-assembly of the blank micelle in an aqueous solvent was tracked through a series of snapshots, as illustrated in Fig. 7, with water beads removed for clarity. Micelle formation is similar across various polymer systems, and PEG10NIP30 was used as a representative sample. At 0 ns, 100 randomly packed PEG-PNIPAM chains were placed in the simulation cell. As the simulation progressed (Figs. 7a & 7b), polymer chains began to aggregate, minimizing unfavorable interactions between water and hydrophobic beads. Self-assembly was triggered by rapid hydrophobic and hydrophilic interactions between PEG and PNIPAM, leading to small aggregates. By 200 ns (Fig. 7c), micelles began to form, and despite continued steps, the micelle remained stable, adopting a rod-like shape by 700 ns, as shown in Fig. 7d. To ensure equilibrium, results were analyzed from the last 200 ns of the simulation span. It is worth noting that, due to the thermo-responsive nature of PNIPAM, the onset of polymer chain self-assembly is influenced by temperature. To investigate this, the kinetics of micelle formation at 300 K and 330 K were analyzed using SASA measurements. As shown in Fig. S.2 of the Supporting Information, elevating the simulation temperature results in faster micelle self-assembly.

### 3.2.2. Structural Evolution of the Micelle

SASA quantifies the surface area of a biomolecule that is accessible to a solvent, providing insights into conformational changes and solubility [58]. The SASA profiles depicted in Fig. 8 reveal a clear trend in self-assembly. Initially high, SASA decreases significantly around 200 ns as polymer chains aggregate, reaching a plateau after 200 ns. At 310 K, all systems self-assemble, with the rate and extent of SASA reduction varying by polymer chain length. In line with the sequential simulation frames shown in Fig. 7, this decrease confirms PNIPAM's transition from dispersion to aggregation, validating the accuracy of the CG model. Self-assembly is driven by minimizing unfavorable interactions between water and hydrophobic domains, with SASA reflecting this process. Above PNIPAM's LCST, the isopropyl groups in PNIPAM become hydrophobic, while the PEG polar residue remains hydrophilic. Micelles shield the PNIPAM moieties from water, promoting self-aggregation. Water molecules initially attached to the polymer chains are expelled by the PEG amines, which play a crucial role in this process, and interact with the PNIPAM amide groups. This allows the hydrophobic isopropyl groups to be securely encapsulated within the micelle, thereby stabilizing its structure. Aggregation occurred faster in NIP25, PEG10NIP20, and PEG10NIP25, likely due to their higher diffusivity [33]. All systems formed by 200 ns, with longer chains taking slightly more time. The total rate of SASA change followed the order: PEG10NIP40 > PEG20NIP25 > PEG10NIP30 > PEG10NIP25 > PEG10NIP20 ~ NIP25, showing how polymer length affects conformation. As SASA shows structural packaging, a system with high molecular packing will display a low SASA [59].

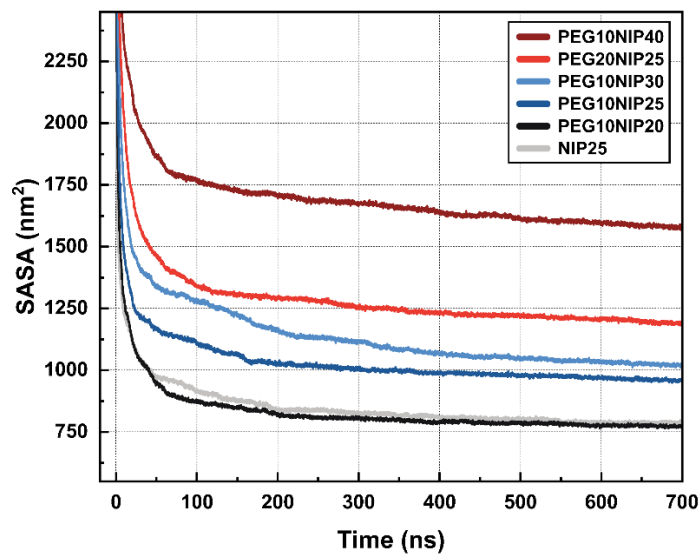

Fig. 8. SASA profile for PEG-PNIPAM blank micelles at 310 K over 700 ns.

As seen in Fig. 8, long-chain systems, like PEG10NIP40, PEG20NIP25, and PEG10NIP30, show increased SASA and solvent exposure due to extended conformations, resulting in larger, less compact micelles. In contrast, shorter chain systems, such as NIP25, PEG10NIP20, and PEG10NIP25, have lower SASA values because of denser structures with fewer solvent-accessible areas. When the PEG length is set at 10 monomers, increasing the PNIPAM chain length (from NIP20 to NIP40) consistently leads to a higher SASA, with the largest increase occurring between NIP30 and NIP40, reflecting the impact of longer PNIPAM segments. Comparing PEG10NIP25 and NIP25 reveals that adding a PEG chain (10 monomers) increases SASA, as the PEG segment introduces additional solvent-accessible surface area due to its hydrophilic properties. If longer PEG chains (e.g., PEG20NIP25) were included, the SASA would likely be even higher, as longer PEG chains tend to increase micelle hydration and size. Although SASA is reduced in all systems, it is less significant in more hydrophilic chains, such as PEG20NIP25, where PEG increases the LCST, enhances fluidity at 310 K, and leads to higher SASA values and weaker packing. The results suggest that PEG-containing systems, particularly those with longer PEG chains, prefer expanded configurations, while neat PNIPAM-based systems are prone to compact structures. Based on Lin et al. [21] findings, water can easily penetrate the PEG corona with higher PEG density and chain length.

### 3.2.3. Structural Characterization

The size and shape of micelles are crucial structural features in controlled drug delivery systems. In MD simulations, Rg is a key indicator of structural formation, offering insights into the compactness, size, shape, and stability of molecules [60]. A lower Rg indicates a more compact and dense structure, while a higher Rg suggests a more extended or dispersed configuration. Table 2 presents the micelle structural parameters, with Rg reflecting conformation and $D_e$, the effective micelle diameter, determined by substituting the values of Rg into Eq. 1 [32,61]:

$$D_e = 2\sqrt{\frac{5}{3}}R_g \tag{1}$$

The time evolution of PEG-PNIPAM micelles (Fig. 9) and their PEG and PNIPAM groups (Figs. S.3a and S.3b) was monitored to demonstrate the impact of chain length on micelles' Rg. Since polymerization degree and molecular weight influence Rg [62,63], with a fixed DP of 100, we analyzed how chain length affects Rg.

Table 2. Rg of copolymer chains, Head (PEG), and Tail (PNIPAM) segments, alongside effective diameter as size indicators

| Sample | Radius of Gyration (nm) | | | Effective Diameter (nm) |
|---|---|---|---|---|
| | Copolymer | Head (PEG) | Tail (PNIPAM) | |
| NIP25 | 6.56±0.58 | - | 6.56±0.58 | 16.93±1.49 |
| PEG10NIP20 | 7.08±1.19 | 7.45±1.37 | 7.16±1.33 | 18.29±3.06 |
| PEG10NIP25 | 7.69±0.56 | 8.03±0.48 | 7.85±0.52 | 19.87±1.45 |
| PEG10NIP30 | 6.46±0.22 | 6.78±0.26 | 6.50±0.23 | 16.69±0.58 |
| PEG10NIP40 | 9.34±0.38 | 9.59±0.30 | 9.31±0.39 | 24.11±0.98 |
| PEG20NIP25 | 8.19±0.63 | 8.82±0.71 | 7.67±0.75 | 21.15±1.63 |

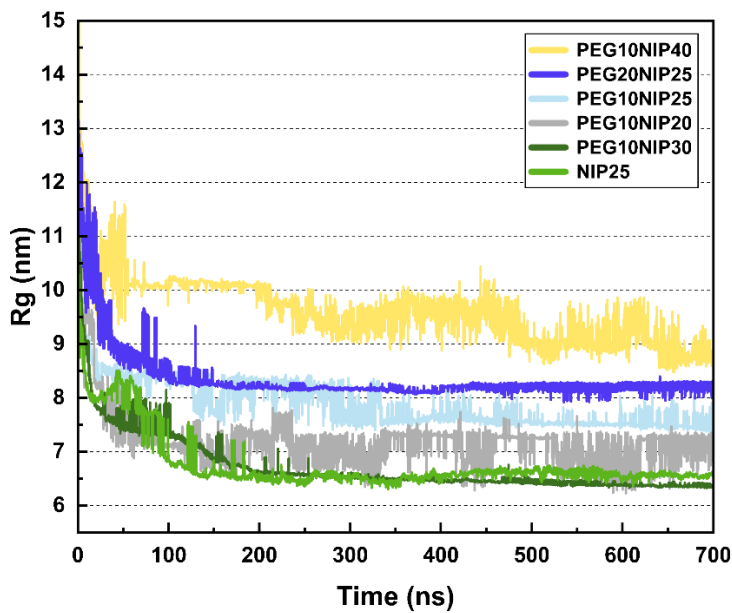

Fig. 9. Time evolution of Rg for copolymer chains in PEG-PNIPAM micelles.

PEG10NIP30 and NIP25 micellar systems exhibit the smallest Rg values, indicating greater compactness and intense aggregation. The lower the Rg, the smaller and more stable the micelles, although this compactness may limit the encapsulation capacity, which can be observed further. NIP25's minimal SASA and Rg suggest a small, compact structure. In contrast, PEG10NIP30 shows moderate SASA along with the smallest Rg values, indicating that polymer chains, especially the hydrophilic PEG segments, are packed more closely as the aggregation size increases. Using short PEG monomers in PEG10NIP30 maintains a balanced structure, providing space for drug encapsulation while preserving micelle stability.

In contrast, PEG10NIP20 and PEG10NIP25 systems show large fluctuations in Rg, without significant decreases, suggesting dynamic self-assembly and potential instability. PEG20NIP25 stabilizes at an Rg of 7.67 ± 0.75 nm, indicating that longer PEG chains increase micelle flexibility and expansion, resulting in higher SASA values while maintaining structural integrity. This flexibility allows the micelle to accommodate a wider range of drug molecules. Longer PEG chains form a dense hydrophilic corona, enhancing water solubility and biocompatibility. Despite being less compact than PEG10NIP30, PEG20NIP25 remains stable under physiological conditions. PEG10NIP40, with the highest Rg (9.31 ± 0.39 nm), forms larger, less compact micelles due to the binding and wrapping effects of increased PNIPAM monomers. Over time, hydrophilic interactions with water cause PEG10NIP40 to expand, reducing compactness and stability. Rg reflects polymer chain aggregation, with stronger adsorption energy pulling chains closer. Higher adsorption energy limits accumulation due to external forces (e.g., shear force, blood circulation), indicating system stability [64]. Unstable Rg values suggest continuous structural changes [65]. The erratic Rg values of PEG10NIP40, PEG10NIP25, and PEG10NIP20 indicate significant structural shifts, while PEG20NIP25, PEG10NIP30, and NIP25 show more compact, stable structures due to intramolecular interactions. Comparing the Rg of PEG (Head) and PNIPAM (Tail) summarized in Table 2 reveals that the head Rg is larger in all systems, allowing calculation of the shell thickness.

Following the analysis of Rg values, the effective diameter ($D_e$) becomes an essential parameter for understanding micelle size and its role in DDSs. This factor is often used as a key indicator for characterizing the size of nanoparticles, as it reflects the overall micellar structure and hydrodynamic behavior in solution [21]. The size characterization of PNIPAM in experimental studies depends significantly on the synthesis route used, as the minimum size of homogeneously cross-linked particles is limited by the polymer's crosslinking time [66]. Previous studies using TEM imaging have shown that PNIPAM-based micelles exhibit a broad size distribution, ranging from less than 50 nm to over 200 nm at both room temperature and 313 K [67]. As shown in Table 3, the effective diameter values align well with both experimental and numerical results, confirming the reliability of our findings. Micelles with smaller sizes, such as those in this study, have been found to be advantageous for DOX loading [68]. Typically, particles smaller than 10 nm are removed from the circulation system, while particles larger than 10 nm can penetrate small capillaries within body tissues [69]. The size of the PEG-PNIPAM micelles increases as the PEG chains are distributed on the surface, as shown in the comparisons made in Table 2 between the NIP25, PEG10NIP25, and PEG20NIP25 models [70].

Table 3. Comparison of effective diameter values with experimental and numerical results from previous studies

| Approach | *De* (nm) | System/Description | Reference |
|----------|-----------|--------------------|-----------|
| CG | 16.69-24.11 | **PEG-PNIPAM** <br> *310 K-with various chain length* | ***Present Work*** |
| CG-Experimental | 30.98-38.73 | **PEG-PNIPAM** <br> *Around phase transition temperature-90 monomer* | [70] |
| AA | 10.33 | **PNIPAM** <br> *313 K- 48 chain- 30 monomer* | [71] |
| Experimental | 10.6-25.6 | **PNIPAM** <br> *298 K-Various molecular weights* | [72] |
| Experimental | 38 | **PNIPAM** <br> *310 K-Hydrodynamic Diameter* | [73] |
| Experimental | 20-100 | **PNIPAM-poly (acrylic acid)** | [74] |
| Experimental | 19.7 nm | **PNIPAM-co-methacrylic acid- co-octadecyl acrylate** <br> *310 K* | [75] |
| Experimental | 34 | **PNIPAM-poly (N-vinylimidazole)** <br> *313 K-Hydrodynamic radius- 44 monomers of NIPAM* | [76] |
| Experimental | 61-117 | **PEG-PNIPAM** <br> *310 K-170 to 270 polymer chain per micelle* | [77] |

Micelles are promising drug delivery carriers due to their ability to self-assemble into diverse structures, including spheres, rods, worms, or disks [78]. To assess copolymer length effects on structural properties, a through shape analysis was performed using ellipticity, PMI, relative shape anisotropy ($K^2$), and eccentricity. Ellipticity ($I_{max}/I_{min}$) measures deviation from sphericity, with 1 indicating a perfect sphere and higher values indicating elongation. The inertia tensor classifies shape: equal diagonal terms suggest a sphere, while differences suggest oblate or prolate forms [79]. $K^2$ quantifies anisotropy, where 0 represents a sphere and 1 indicates maximum elongation [32]. Eccentricity measures elongation, with 0.0 denoting a perfect sphere [80].

The results summarized in Table 4 show an eccentricity between 0.66 and 0.77, indicating a significant deviation from sphericity and suggesting an ellipsoidal micellar structure. The PMI ratios align with eccentricity and ellipticity values, confirming an elliptical morphology for PEG-b-PNIPAM micelles, consistent with previous PEG-PNIPAM studies [37]. Morphological parameters confirm the ellipsoidal aggregate observed in dynamic snapshots illustrated in Fig. 7. Non-spherical micelles, such as rods and worms, likely result from overpacking of hydrophobic moieties within the core, minimizing the aggregation risk [32]. Notably, rod-like micelles deliver drugs to tumors more effectively than spherical ones [81]. Also, larger rod-like micelles better shield and protect drug molecules and internal structures [32]. Wormlike micelles exhibit higher SASA (Table 4) due to their elongated structure, which increases solvent exposure. In rod-shaped PEG-PNIPAM micelles, water molecules more easily access PNIPAM groups than in spherical counterparts, thinning the middle regions. NIP25 and PEG10NIP20 form elongated, moderately anisotropic structures, while PEG10NIP25 and PEG10NIP30 show the highest elongation. As PNIPAM segments increase (PEG10NIP20, 25, 30), micelles become unstable and elongated, while PEG10NIP40 regains symmetry and resembles a sphere. PEG20NIP25, with longer PEG chains, forms the most stable and symmetric micelles. These findings suggest that polymer chain length affects micellar shape, stability, and drug delivery potential.

Table 4. Shape descriptors and surface area of the micelles

| Sample | Ellipticity | $K^2$ | Eccentricity | I1/I2 | I1/I3 | I2/I3 | SASA (nm$^2$) |
|---|---|---|---|---|---|---|---|
| NIP 25 | 2.52±0.01 | 0.06 | 0.78 | 0.51±0.03 | 0.40 | 0.78±0.04 | 771.85±35.83 |
| PEG10NIP20 | 2.27±0.09 | 0.05 | 0.75±0.01 | 0.62 | 0.44±0.02 | 0.71±0.03 | 794.42±21.20 |
| PEG10NIP25 | 2.66±0.35 | 0.06±0.01 | 0.78±0.03 | 0.46±0.07 | 0.39±0.05 | 0.869±0.04 | 975.06±32.49 |
| PEG10NIP30 | 2.50±0.10 | 0.06 | 0.77±0.01 | 0.53±0.02 | 0.40±0.01 | 0.75 | 1079.29±3.07 |
| PEG10NIP40 | 1.80±0.19 | 0.03 | 0.66±0.03 | 0.72±0.10 | 0.56±0.04 | 0.80±0.09 | 1641.62±41.06 |
| PEG20NIP25 | 1.82±0.14 | 0.03 | 0.66±0.04 | 0.66±0.10 | 0.56±0.05 | 0.85±0.05 | 1237.11±37.80 |

### 3.2.4. Micellar Morphology and Spatial Distribution

A crucial aspect of micelle or hydrogel formation from polymeric blocks is understanding their structural partitioning and spatial distribution. Insight into the internal structure of PEG-PNIPAM micelles can be gained by examining the radial density of their components relative to the micelle's center of mass (COM). Number density measures the absolute density of a segment at a given distance, $r$. Unlike normalized RDF values, it retains physical units, allowing direct density comparison. Higher peaks in the radial density profiles in Fig. 10 highlight regions with greater molecular concentration. The highest PNIPAM distribution is in the micelle core, while PEG is mainly found in the shell, with its distribution gradually decreasing outward, indicating a diffuse outer layer. The water content in the core is low but increases toward the shell, with peak density in the outer layers, suggesting water infiltration. Water distribution around PEG is much higher than around PNIPAM, reinforcing the idea that PEG forms the outer shell, more exposed to water, while PNIPAM is tightly packed inside the hydrophobic core [37].

Micelle density profiles further confirm the spatial arrangement of PEG, PNIPAM, and water molecules (see Fig. S.4 in the Supporting Information). The presence of PNIPAM beyond the core layer implies that the PEG-PNIPAM micelles adopt an ellipsoidal shape. The RDF curve of PEG residues in PEG20NIP25 overlaps with PNIPAM atoms. The hydrophilic PEG occupies the outer corona, most exposed to the solvent, while lower peaks in other systems indicate regions with entangled micellar structures. In PEG10NIP40, where the PEG/PNIPAM ratio is low, some PEG groups reside deeper inside the core, suggesting that a few PNIPAM units are arranged with PEG groups pointing toward the micelle core. NIP25, PEG10NIP25, and PEG10NIP30 form water-free cores, whereas other systems have a small amount of water in the core. The RDF results indicate that incorporating PNIPAM or PEG into the PEG-PNIPAM micellar systems does not change their overall structure. Instead, the PEG:PNIPAM ratio plays a significant role in core solvation, polymer entanglement, and interface between the hydrophilic and hydrophobic domains.

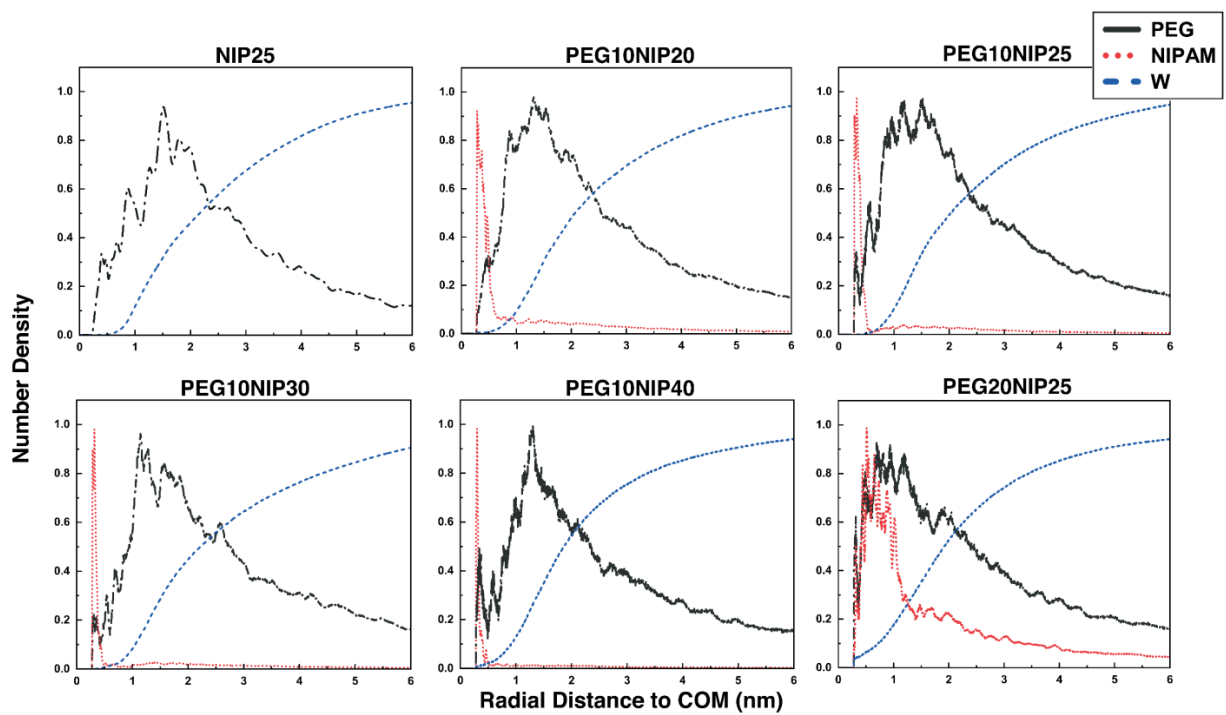

Fig. 10. Radial density profiles and spatial distribution of PEG, PNIPAM, and water molecules within the micelles.

### 3.2.5. Solvent Coordination and Segmental Interactions

RDF analysis provides insight into local structure, with the first peak in $g(r)$ indicating the closest distance and peak height reflecting interaction probability. RDFs between CG sites reveal short-range molecular organization, while RDFs between micelle heads and tails relative to water demonstrate comparable molecular interactions. RDFs from CG simulations of a single PEG-PNIPAM chain (Fig. S.5 in the Supporting Information) closely align with atomistic simulations by Rezaeisadat et al. [37]. These RDFs reveal that water distribution around the hydrophilic PEG segment is higher than that of the hydrophobic PNIPAM part. Thus, PEG is considered the micellar head, and PNIPAM, the tail.

In Figs. 11a & 11b, the hydration behavior of the PEG and PNIPAM segments of the copolymer chains is compared to assess their water affinity. Next, Fig. 12 offers a broader perspective by illustrating the hydration of the entire copolymer chain in the self-assembled structures (polymer-water RDF). Notably, insets have been added to all the graphs to facilitate peak comparison. RDFs from the final 200 ns of simulation in these figures show distinct peaks at $r = 0.26$–$0.27$ nm for PNIPAM and at $r = 0.35$ nm for PEG, indicating specific interactions between water and the segments. The peak positions across all six systems were similar, but the height of the first peak gradually decreased, suggesting a reduction in bond formation and the development of well-defined water shells around nonpolar and polar beads. It is worth mentioning that due to the homogeneous structure of the developed nanomicelle moieties, the RDF value increased at a certain distance before sharply decreasing [82]. The primary difference between the RDF curves lies in their intensities and positions. The PNIPAM (tail) shows higher and closer radial distribution patterns around water molecules compared to the PEG (head). Despite PEG's hydrophilic nature, the water density around the hydrophobic PNIPAM domains (C1, Nda, and C3) is higher than that of the PEG regions (N0 and P2). This is attributed to the lower density of PEG, which allows entanglement and fewer interactions with water molecules. Another contributing factor is the loose structure of the polymeric micelle, which allows water beads to penetrate the hydrophobic layers. PEG shows more peaks than PNIPAM, indicating more hydration shells and hydrogen bonding between PEG's ether groups and water. PEG's flexibility facilitates these interactions along the entire polymer chain.

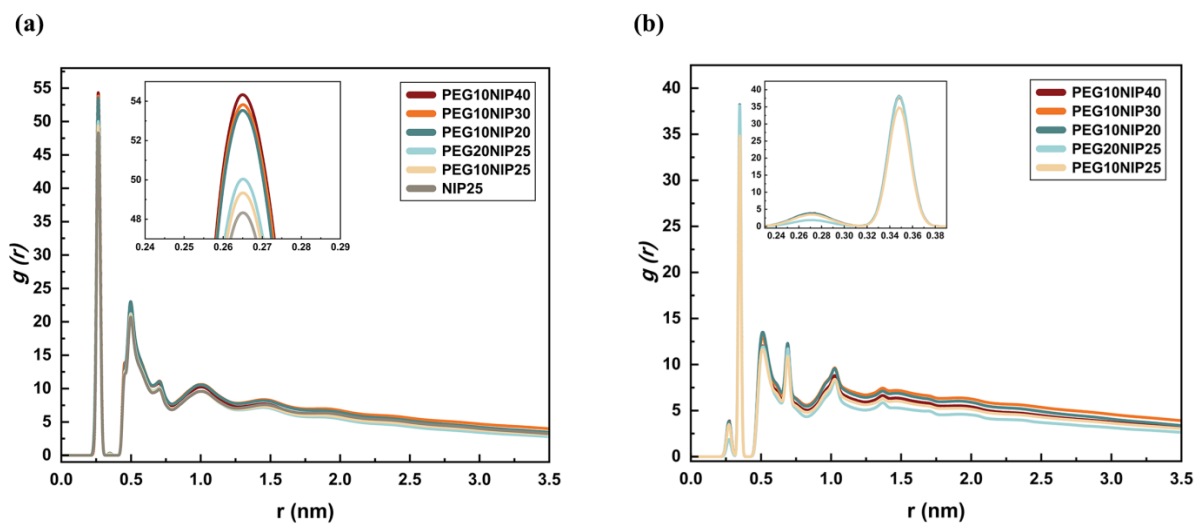

Fig. 11. RDF for (a) PNIPAM and (b) PEG segments of CG PEG-PNIPAM blank micelles and water.

The PNIPAM-water RDF depicted in Fig. 11a highlights how water molecules interact with the hydrophobic PNIPAM segments. Since PNIPAM forms the core, this plot reflects the extent of micelle core hydration, which is essential for micelle stability and hydrophilic/hydrophobic balance in DDSs. PEG10NIP40 exhibits the highest sharp first peak, indicating strong hydration near the PNIPAM segments due to its large PNIPAM chain. The prominence of this peak suggests that the large PNIPAM enhances water accessibility to the core, correlating with the system's hydrophilicity and larger SASA values. Systems with higher hydration peaks, like PEG10NIP40, show more expanded structures. This adsorption capacity gradually decreases as the PNIPAM chain length shortens to PEG10NIP30. This is followed by a slightly lower peak, indicating moderate hydration as PNIPAM domains become more densely packed. PEG10NIP20 and PEG10NIP25 exhibit progressively lower hydration levels, with PEG10NIP25 having a more compact structure and lower SASA. In contrast, PEG10NIP20, with a looser structure, allows more water access. The trends in PEG10NIP20, PEG10NIP25, and PEG10NIP30 are attributed to their intermediate PNIPAM content, balancing hydrophobic and hydrophilic interactions [83]. Due to the compact micelle core, PEG10NIP25 and NIP25 show the lowest hydration peaks, indicating hydrophobic regions and weaker water interactions. A comparison of NIP25, PEG10NIP25, and PEG20NIP25 reveals that adding PEG increases water accessibility in the core region. The peak at ~0.5 nm, which corresponds to the second hydration shell, is most prominent in the PEG10NIP40, PEG10NIP30, and PEG10NIP20 models, indicating extended hydration structures.

This larger shell reflects a more hydrophilic micelle, driven by PNIPAM dominance in PEG10NIP40 and PEG10NIP30 and a looser structure in PEG10NIP20. For other systems, the subsequent peaks decrease in intensity, suggesting tighter micelle packing.

Using RDF analysis, the hydration behavior and water structuring around the hydrophilic PEG head are revealed (Fig. 11b). The first prominent peak, where water molecules interact directly with PEG heads, shows similar hydration in PEG10NIP40, PEG10NIP30, PEG10NIP20, and PEG20NIP25 micelles. The PEG heads in these systems are equally accessible to water molecules, suggesting that PNIPAM composition has little effect on the direct water interaction around PEG and the entangled structure of these segments. The lower peak for PEG10NIP25 suggests reduced PEG-water interaction due to its more compact structure, which limits PEG exposure to surrounding water. Increasing the PEG segment from PEG10 to PEG20 improves water accessibility, possibly due to the shielding effect of the longer PEG chains. The second and third peaks are most prominent in PEG10NIP30 and PEG10NIP20, followed by PEG10NIP40, indicating enhanced long-range water ordering. PEG chains exposed to more water and expanded micelle structures facilitate better water penetration. The lowest second and third peaks in PEG10NIP25 reflect reduced water structuring due to its compact structure, limiting water penetration into the hydrophilic region.

To analyze the hydration environment of PEG-PNIPAM in their self-assembled nanostructures, RDFs were also computed between the copolymer chains and water beads (Fig. 12). The profiles show similar patterns, with the first peak at 0.5 nm and the second at 1 nm, reflecting well-defined hydration near the micelle surface, with uniform water distribution in the solution phase and fewer water molecules surrounding the micellar core. The peak positions remain unchanged with varying molecular weight, but peak intensities vary significantly. The PEG20NIP25 system shows the highest peak, demonstrating stronger hydration and higher hydrophilicity, likely due to its larger PEG segment. This enhances micelle hydrophilicity and increases water accessibility. The consistently high RDF values of the cyan curve beyond the first peak indicate prolonged hydration in the micelle's outer regions, which corresponds to its larger Rg and SASA. The PEG10NIP40 system has a slightly lower RDF peak than PEG20NIP25, indicating moderate hydration levels. Despite its larger PNIPAM length, good hydration is observed because of the longer chains. Larger Rg (PEG20NIP25, PEG10NIP40) correlates with higher RDF values because of the increased surface area for water interaction. The PEG10NIP25 and PEG10NIP20 curves show intermediate hydration behavior between PEG10NIP40 and PEG10NIP30, reflecting a balanced contribution of PEG and PNIPAM to hydration. The moderate first peak height suggests adequate water interaction, but the flatter tail points to reduced long-range hydration compared to PEG20NIP25. This stems from the small PNIPAM core and limited PEG coverage, which lowers water interaction. The PEG10NIP30 curve shows slightly lower RDF peaks than PEG10NIP20, indicating less hydration and fewer additional, broader peaks. The compact structure of PEG10NIP30, reflected by its smaller Rg, results in reduced water accessibility and lower RDF values.

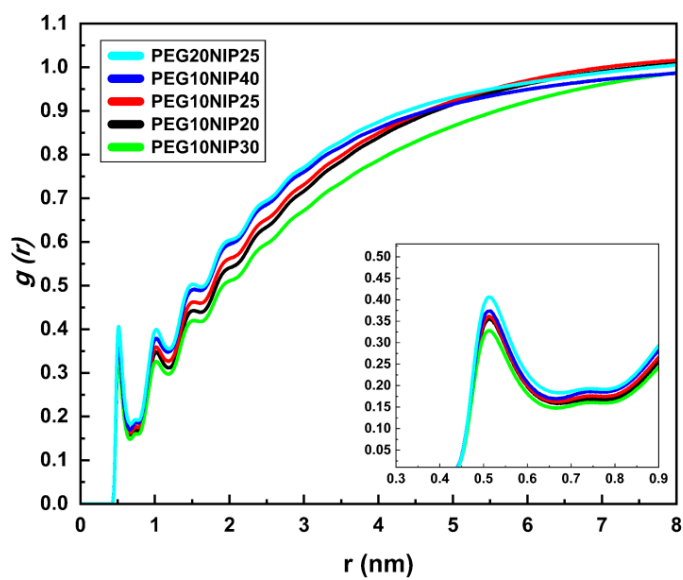

Fig. 12. RDF of copolymer chains and water in blank PEG-PNIPAM micelles.

Polymer chains rapidly oriented toward the hydrophilic PEG shell and wrapped around PNIPAM segments during the spontaneous aggregation process. Since PNIPAM is hydrophobic and PEG is hydrophilic, polymeric micelles form a semi-core-shell structure, with the hydrophilic shell stabilizing and dispersing the micelles. This micelle aggregate differs from other "core-shell" models, which generally consist of hydrophobic amphiphilic molecules with polar head groups in the outer shell. We propose that this difference arises from varying hydrophobicities and corresponding ratios. Overall, the findings suggest that PEG-PNIPAM systems can self-assemble into elongated, ellipsoidal polymeric micelles within 200 ns, with self-assembly driven by hydrophobic interactions.

To complement our chain length analysis and address additional factors relevant to *in vivo* conditions, we include here a discussion on how environmental variables can influence the behavior of PEG-PNIPAM micellar systems. While our study maintains constant conditions to isolate the effect of chain length, it is important to acknowledge the broader physicochemical context. As discussed in the literature, salt concentration, pH, and serum proteins can significantly affect the LCST and PEG hydration. The impact of added salts depends on their nature; cosmotropic salts tend to increase LCST, while chaotropic salts decrease it [84,85]. Variations in pH influence the charge distribution and hydrogen bonding within PNIPAM chains [86]. In pure water or weak acidic/basic environments, hydrogen bonds formed by the isopropyl-substituted amide groups limit PNIPAM swelling. However, under strong acidic or basic conditions, these hydrogen bonds break, leading to partial ionization and chain expansion due to enhanced interactions with water [87]. Once PEG enters the bloodstream, it rapidly encounters serum proteins that form a protein corona, a key factor affecting its circulation time. These protein interactions can alter the LCST, influence PEG behavior in solution, and modulate hydration depending on the protein type and concentration [85,88]. Despite these influences, the present study specifically investigates chain length as a primary factor in self-assembly and drug encapsulation, using pure, neutral aqueous solutions at a constant physiological temperature. Holding other variables constant allows for direct comparison across simulations and ensures a clear interpretation of chain length effects.

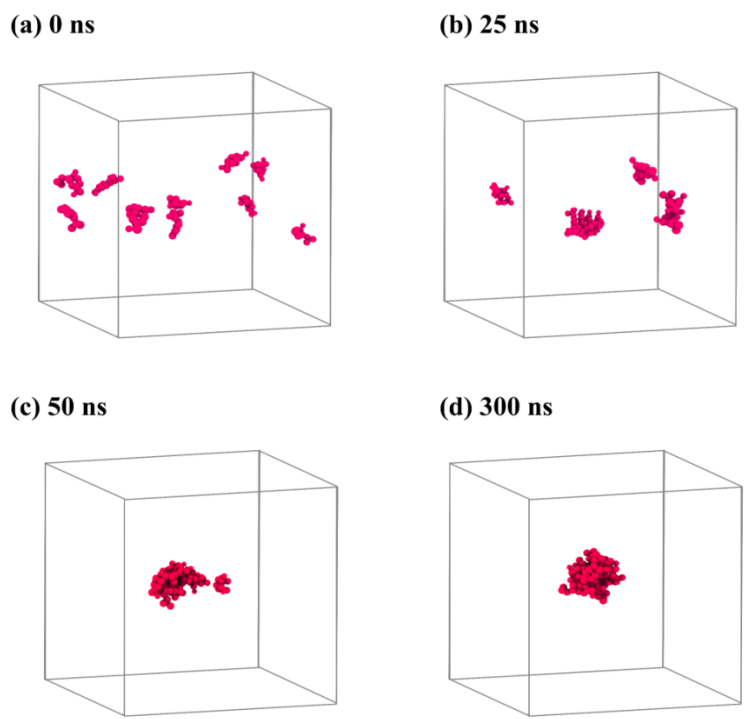

Fig. 13. A series of simulation snapshots showing DOX aggregation behavior in aqueous solvent over time: (a) 0 ns, (b) 25 ns, (c) 50 ns, and (d) 300 ns.

### 3.3. DOX Encapsulation and Drug-Loaded PEG-PNIPAM Micelles

A major challenge in pharmaceuticals is the poor solubility of therapeutic drugs. This study aimed to optimize PEG-PNIPAM micelles for DOX encapsulation. To tackle this issue, systems with 100 copolymer chains (10 wt.%) and 1 wt.% DOX were simulated for 700 ns. The simulation initially examined free DOX aggregation, followed by its encapsulation within the polymeric micelles. Drug solubility, localization, and system stability were assessed using simulation snapshots, SASA, Rg, interaction energy, RDF, and MSD.

#### 3.3.1. Pharmacokinetics and Molecular Insights into Free DOX Aggregation

Pharmacokinetics (PK) describes a drug's journey through the body, incorporating the processes of Absorption, Distribution, Metabolism, and Excretion (ADME). The ADME studies of DOX are provided in the Supporting Information. Drug solubility plays a critical role in DDSs. To track DOX solubility behavior, a system with 10 DOX molecules at 1 wt.% was modeled. Figure 13 depicts snapshots of DOX aggregation at 0 ns, 25 ns, 50 ns, and 300 ns, indicating that DOX molecules aggregate, as do other amphiphilic molecules, to minimize unfavorable water-hydrophobic interactions. For a quantitative investigation, we considered three key parameters: SASA, Rg, and RDF between water molecules and DOX. Figure S.6 in Supporting Information provides the corresponding results for these parameters. SASA values decrease during aggregation, plateauing at 35 nm², representing the fully aggregated state. Rg values show an aggregation trend, reaching 1 nm after 100 ns of simulation. The RDF curve indicates the preferred distance of water around the molecule, showing the formation of a hydration shell around the structure.

### 3.3.2. Dynamic Encapsulation Snapshots of Drug-Loaded PEG-PNIPAM Micelles

Time-lapsed snapshots of drug-loaded PEG-PNIPAM micelle formation (using PEG20NIP25 as a representative sample) are shown in Fig. 14. The encapsulation process reveals similar self-assembly patterns across different systems. Initially, DOX molecules and PEG-PNIPAM copolymer chains are dispersed in the simulation box. By the end of the simulation (700 ns), the chains are highly aggregated, and DOX spontaneously solubilizes into the micelle, resulting in encapsulation. Finally, hydrophobic interactions cause polymers to accumulate, trapping DOX within the formed ellipsoidal micelle.

Despite increasing the simulation time, the micelle size and morphology remained largely unchanged, indicating that the system had reached equilibrium and the drug-loaded micelles were stable. Due to interactions between the drug and solvent molecules, the drugs predominantly reside on the contact surface of the micelles with the solution. The resulting drug-loaded micelles displayed an ellipsoidal shape, similar to the blank micelles. These findings suggest that DOX co-assembles with the copolymer chains without altering the self-assembly capability of the micelle. Additionally, DOX molecules were randomly distributed within the micelle. Comparing the aggregation results for free DOX (see Figure S.6 in Supporting Information), it is deduced that PEG-PNIPAM micelles effectively prevent DOX aggregation, enhancing its solubility in biological media.

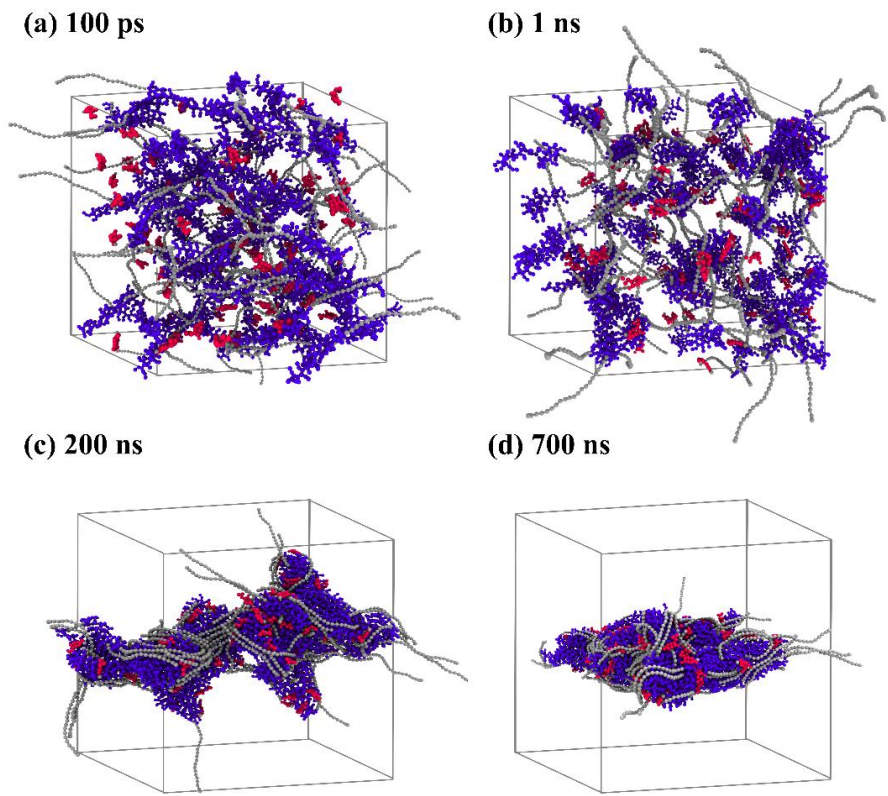

Fig. 14. A series of simulation snapshots of the drug-loading process for DOX in PEG-PNIPAM micelles in aqueous solvent over time: (a) 100 ps, (b) 1 ns, (c) 200 ns, and (d) 700 ns.

### 3.3.3. Hydration and Solvent Exposure of DOX and Drug-Loaded Micelles

The surface contact area of DOX molecules and drug-loaded micelles with water was analyzed using SASA, a key indicator of nanocarrier stability. The stable SASA profile of encapsulated DOX over time (Fig. 15a) reveals the transition from free DOX to its encapsulated form. The dominance of encapsulation over aggregation kinetics is crucial for inhibiting drug aggregation and achieving high drug-loading efficiency [89]. Faster encapsulation than aggregation of DOX at the same concentration indicates that PEG-PNIPAM micelles can effectively solubilize the drug.

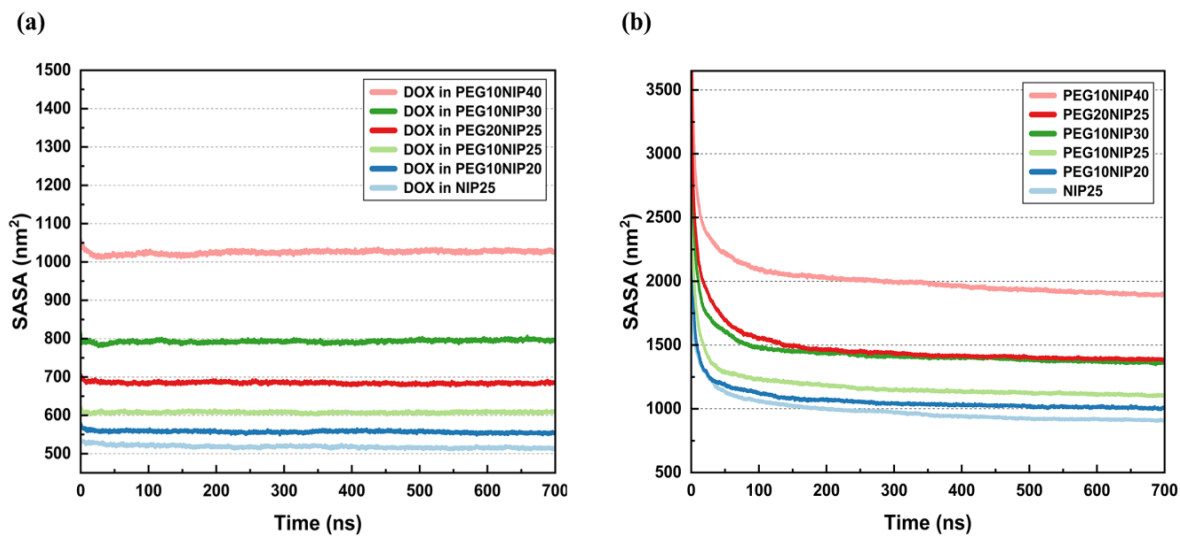

Fig.15. SASA of (a) encapsulated DOX and (b) drug-loaded micelles over simulation time.

Due to the tendency of DOX molecules to bind to the shell, micelles containing DOX are expected to present a larger surface area accessible to water. Indeed, the surface area of drug-loaded PEG-PNIPAM micelles exceeds that of blank ones (see Fig. 15b & Fig. 8). Consequently, DOX adsorption promotes micelle coalescence, meaning the drug is attached to the micelle shell rather than its surface [90]. This suggests that the drug avoids sudden release. Although contact area analysis reveals the intensity of adsorption/repulsion, it does not distinguish the underlying mechanisms. In the NIP25 model, the DOX SASA mean value is the lowest among various polymer chain lengths, while the highest is observed with the PEG10NIP40 system. This trend is consistent for drug-loaded micelles, where longer polymer chains result in higher SASA values. The PEG10NIP40 system provides the largest surface area for DOX, making it more suitable for encapsulation. Conversely, systems with shorter polymer chains exhibited lower SASA values, suggesting better adsorption efficiency, thus offering more protection to DOX from water.

In both blank and drug-loaded micelles, SASA increases with longer chain segments, except in drug-loaded PEG20NIP25 and PEG10NIP30. Notably, in PEG10NIP30, the addition of DOX disrupts the hydrophilic/hydrophobic balance, causing the micelle to expand. In contrast, PEG20NIP25 exhibited a lower DOX SASA value, suggesting that DOX molecules are located deeper within the micelle's inner layers, making it a more effective adsorbent. These findings emphasize the significance of copolymer block length and ratio on micellar stability. Additionally, the degree of flexibility or rigidity of the copolymer blocks influences drug distribution within the micellar shell. Rigid PNIPAM chains prevent drug aggregation in the hydrophobic core, while flexible PEG domains allow mobility, promoting drug aggregation at the micelle's center.

### 3.3.4. Structural Effects of DOX Loading on Drug-Loaded Micelles

Drug-loaded micelles form when the system reaches a steady state and the Rg of micelles stabilizes (Fig. 16a). The Rg values follow this sequence: NIP25 < PEG10NIP25 < PEG10NIP20 < PEG20NIP25 < PEG10NIP30 < PEG10NIP40. Generally, micelles with shorter copolymer chains exhibit lower Rg values, consistent with previous findings that shorter polymer chains lead to more stable complexes. For systems NIP25, PEG10NIP20, and PEG20NIP25, the Rg of DOX-loaded micelles increases by 1–5% compared to blank micelles, indicating a slight increase in micellar size with 1% drug concentration. This result aligns with previous studies [32,91]. In contrast, the Rg of PEG10NIP30 and PEG10NIP40 increases by 40% and 6.85%, respectively, with gradual increases observed over the last 200 ns. These systems tend to form unstable aggregates, leading to an overall Rg increase. Therefore, PEG10NIP30 and PEG10NIP40 do not aggregate into stable structures but rather form dynamic aggregates. This behavior has been previously observed in PNIPAM-Curcumin micelles [92], though longer simulation times are needed to confirm these findings.

The Rg values for the PEG10NIP25 system with adsorbed DOX are lower than those for the blank micelle, suggesting that DOX binding enhances micelle stability and compactness. This trend, also reflected in SASA analysis, indicates that DOX binding leads to a more stable structure with minimal impact on particle size. Similar trends are observed in the Rg of encapsulated DOX over the last 200 ns shown in Fig. 16b. Generally, average Rg reflects the drug density within aggregates and indirectly indicates drug-loading stability. Smaller Rg values correlate with higher drug density and better dynamic stability. Systems NIP25, PEG10NIP25, and PEG10NIP20 have lower Rg values than PEG20NIP25, PEG10NIP30, and PEG10NIP40, indicating better DOX stability in models with shorter polymer chains, as well as PEG20NIP25. Notably, the PEG ratio significantly affects DOX loading performance at a 1% feeding concentration. More rigid PNIPAM chains restrict drug movement during self-assembly, preventing drug aggregation in hydrophobic micelle cores.

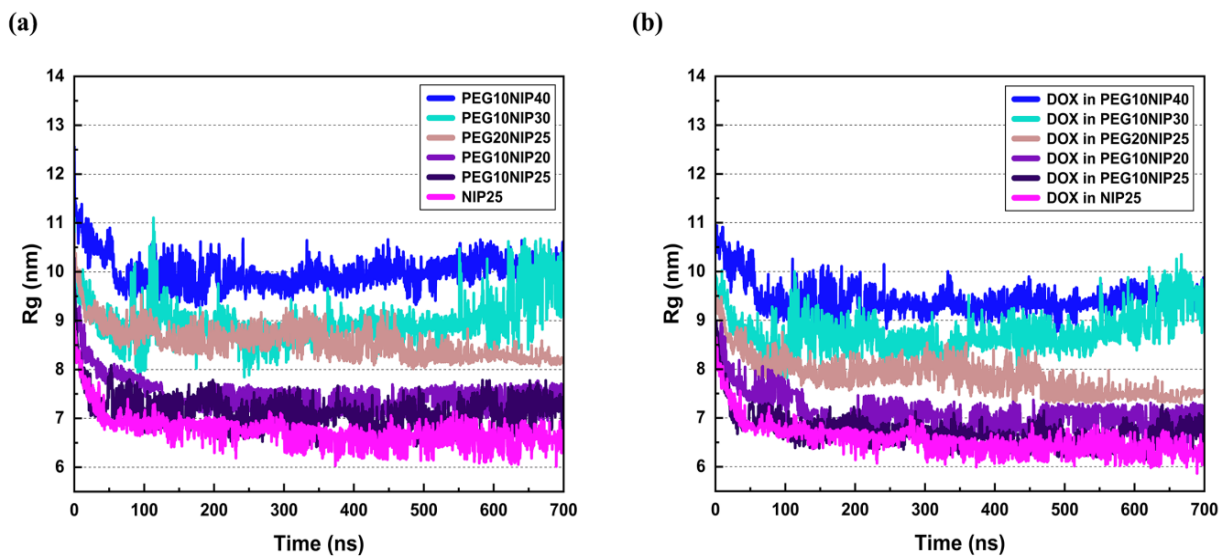

Fig. 16. Rg values of (a) drug-loaded micelles and (b) encapsulated DOX over the simulation time.

Considering changes in micellar diameter during the encapsulation process, we also examine the shape of the micelles using shape descriptors. Typically, PNIPAM-based micelles are elliptical rather than spherical. For an effective DDS, the micelle should retain its shape post-loading. Previous research has shown that PEG-PNIPAM micelles maintain their elliptical shape despite changes in Rg during encapsulation [37]. A comparison of eccentricity and ellipticity for drug-loaded micelles (Table 5) revealed that the drug-loaded micelles maintain the same ellipsoidal shape as blank micelles, with slight changes in shape parameters. For example, PEG10NIP40, PEG20NIP25, and PEG10NIP25 showed a 1-6% change in their shape. PEG10NIP40 system tends to adopt a more ellipsoidal shape (4%), while PEG20NIP25 (0.3%) and PEG10NIP25 (6%) tend toward more spherical shapes. The most pronounced changes were observed in systems NIP25, PEG10NIP30, and PEG10NIP20, with ellipticity changes of 9%, 16%, and 22%, respectively.

The addition of DOX to NIP25 micelles resulted in a more ellipsoidal structure. The most significant change in ellipticity was observed in PEG10NIP20 micelles, where DOX incorporation caused a shift toward a more ellipsoidal shape. This was confirmed by SASA and Rg values, which showed that DOX-loaded NIP25 and PEG10NIP20 micelles have a compact and smaller size. In contrast, PEG10NIP30 micelles became more spherical, with their expansion occurring in the directions of sphericity. Additionally, the $K^2$ values indicated a transition from spherical to rod-like anisotropic structures in PEG10NIP40 (1%), NIP25 (9.5%), and PEG10NIP20 (36%) systems, suggesting the formation of more elongated micelles. This change is typically linked to enhanced drug shielding in rod-like micelles [32]. On the other hand, a decrease in $K^2$ was observed in the PEG20NIP25 (1%), PEG10NIP25 (~8%), and PEG10NIP30 (~29%) models, signifying the transition from rod-like blank micelles to quasi-spherical drug-loaded counterparts. Spherical geometry is considered ideal for particles moving through the body, as cylindrical micelles are often long and vesicles are large, which can complicate their transport [4]. This study highlights that short polymer chains are more prone to shape changes upon DOX loading, while longer polymer chains undergo minimal shape alterations. Consequently, at a fixed polymer and drug concentration, the final shape and uniformity of the micellar aggregates can be controlled, likely through a mechanism similar to that governing the influence of hydrophobic chain length on micelle structure [32]. Furthermore, when the number of polymers in the system is fixed, the presence of non-spherical structures in the micelles is largely influenced by the incorporation of hydrophobic drug molecules. Both visual inspection and quantitative analysis revealed that DOX molecules caused the formation of non-spherical and quasi-spherical micelles, with only slight effects on the final particle size (see Table 2 and Table 5).

Table 5. Shape and size descriptors for drug-loaded micelles and Rg for encapsulated drugs

| Sample | Ellipticity | $K^2$ | Eccentricity | Rg micelle (nm) | $D_e$ micelle (nm) | Rg Drug (nm) |
|---|---|---|---|---|---|---|
| NIP 25 | 2.75±0.61 | 0.06±0.02 | 0.77±0.05 | 6.63±0.71 | 17.13±1.84 | 6.44±0.57 |
| PEG10NIP20 | 2.78±0.22 | 0.07±0.01 | 0.80±0.02 | 7.41±0.58 | 19.11±1.49 | 6.99±0.58 |
| PEG10NIP25 | 2.52±0.12 | 0.06 | 0.77±0.01 | 6.92±0.38 | 17.87±0.98 | 6.62±0.34 |
| PEG10NIP30 | 2.09±0.20 | 0.04±0.01 | 0.72±0.03 | 9.04±0.72 | 23.33±1.87 | 8.78±0.66 |
| PEG10NIP40 | 1.86±0.15 | 0.03 | 0.67±0.03 | 9.98±0.56 | 25.77±1.45 | 9.39±0.38 |
| PEG20NIP25 | 1.81±0.06 | 0.03 | 0.67±0.01 | 8.46±0.88 | 21.83±2.28 | 7.73±0.83 |

It is important to note that higher drug loading concentrations can significantly influence the morphology, stability, and release behavior of PEG–PNIPAM micelles. Elevated payloads may lead to micelle growth, shape deformation, or even aggregation, particularly when strong hydrophobic interactions between drug molecules are present [93]. While increased loading may enhance the initial release due to a steeper concentration gradient, it can also hinder sustained release as a result of denser drug packing within the micelle core [94]. Excessive loading may further result in the uncontrolled release of large drug quantities, increasing the risk of toxicity [93]. To examine the effects of drug concentration, we conducted MD simulations of 100 PEG20NIP25 chains with 3 wt.% and 5 wt.% DOX, focusing on their self-assembly behavior. The observed increase in Rg to $7.59 \pm 0.36$ nm at 3 wt.% and $8.67 \pm 0.57$ nm at 5 wt.% loading, along with corresponding changes in SASA values to $1762.36 \pm 11.46$ nm² and $2397.50 \pm 12.88$ nm², respectively, indicate micelle size enlargement at higher drug loadings. All shape descriptors consistently showed a shift toward a more ellipsoidal morphology. Specifically, $K^2$ values increased to $0.69 \pm 0.06$ and $0.80 \pm 0.08$, while eccentricity rose to 0.99 and 1.00 for 3 wt.% and 5 wt.% DOX, respectively.

### 3.3.5. Radial Localization and Spatial Distribution of Encapsulated Doxorubicin

Drug location within pharmaceutical carriers is critical for micellar drug delivery, influencing drug loading, formulation stability, and release. Figure 17 presents the radial density plots, showing the density of atoms in shells around the COM. These profiles confirm that PNIPAM groups form tightly packed cores with little to no water, while PEG chains are highly hydrated and located on the micelle surface. Entrapped DOX molecules are primarily distributed between the shell and the core-shell interface, with most drug molecules found in the PEG region. Due to the hydrophobicity of DOX, most drug molecules reside between the PEG and PNIPAM blocks. On average, the drug's location spans 3.91 to 12.18 Å from the COM of the PEG-PNIPAM structure. This suggests that DOX is primarily distributed around the polymer shell but is spread throughout the micelle.

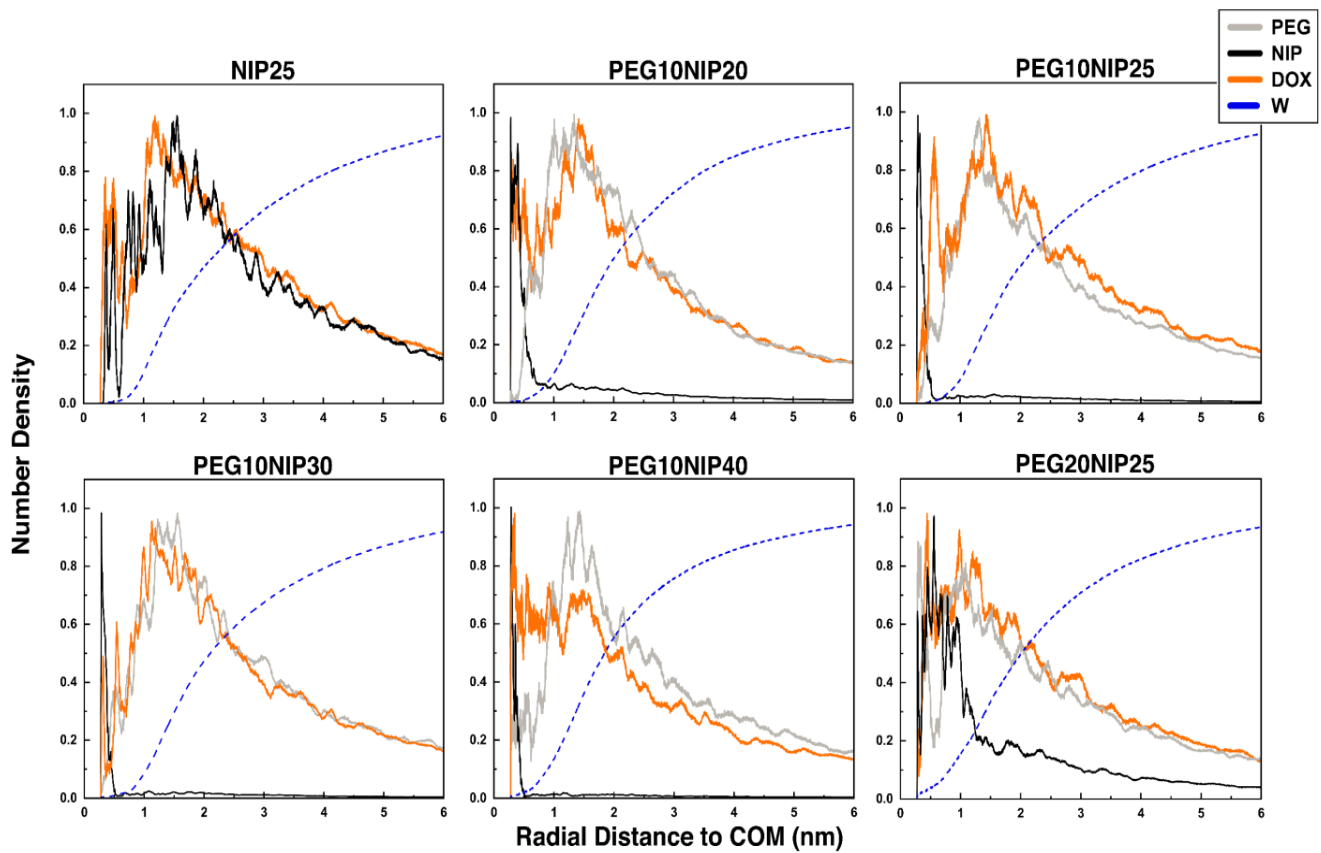

Fig. 17. Radial density profiles and spatial distribution of PEG, PNIPAM, DOX, and water molecules within the drug-loaded micelles.

The broad distribution of drug molecules within the micelles indicates unstable solubilization, potentially leading to burst release or drug precipitation. Density profiles for micelles, drug, and water molecules shown in Fig. S.7 support this. However, entering the hydrophobic core requires overcoming significant steric hindrance [95], and only a small number of DOX molecules are found in the core. In line with previous studies, at low drug concentrations, DOX primarily resides in the PEG layers of the micelle due to resistance when attempting to access the core [96]. In smaller systems (e.g., NIP25, PEG10NIP25, and PEG10NIP30), DOX is confined mostly to the shell region, with limited diffusion into the core. In more flexible systems (e.g., PEG10NIP20, PEG10NIP40, and PEG20NIP25), DOX is more likely to diffuse into the core region. Three factors may explain the limited diffusion of DOX into the PEG-PNIPAM micelle: (1) repulsive interactions with the core, (2) attractive interactions with the shell, and (3) attractive interactions with water. Further analysis through RDF, energy, and diffusion could clarify these mechanisms.

### *3.3.6. Thermodynamic Insights into DOX Encapsulation in PEG-PNIPAM Micelles*

The self-assembly of surfactants is inherently linked to energy changes, reflecting the physical and chemical nature of adsorbate molecules and their intermolecular interactions. Accordingly, interaction energies between various components were analyzed to explore the factors affecting drug solubilization (Fig. 18). Negative values indicate a spontaneous adsorption process. To gain deeper insight into the drug solubilization process, we monitored the variation in interaction energies between DOX-DOX, DOX-Water (DOX-W), DOX-PEG-PNIPAM (DOX-Micelle), and PEG-PNIPAM-Water (Micelle-W). The energy for DOX-DOX remained stable throughout the encapsulation process, indicating monomeric solubilization of DOX [33]. In contrast, the DOX-Water and DOX-Micelle interactions showed opposite trends; DOX-Micelle interactions became more favorable, stabilizing the drug carrier system [33,37]. The increase in DOX-Water interaction energy during encapsulation suggests a hydrophobic solubilization process, although it remains negative, demonstrating some degree of interaction with water.

.

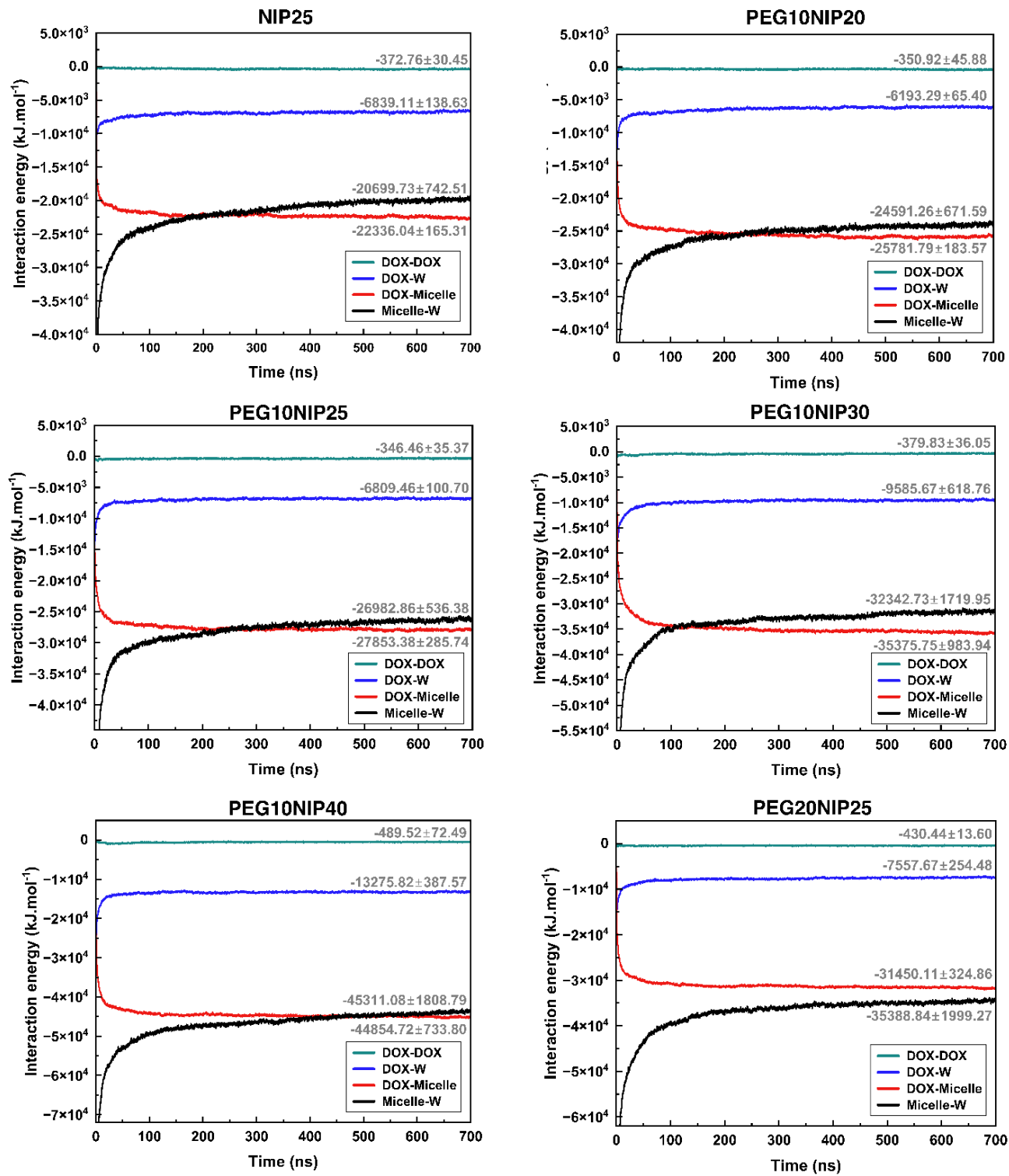

Fig. 18. Interaction energies between various components of the systems, including DOX-DOX, DOX-W, DOX-Micelle, and Micelle-W.

The profiles also show a transition from Drug-Micelle to DOX-W interactions as encapsulation progresses. Notably, the absolute change in DOX-Micelle interaction energy is larger than that of DOX-W, suggesting that DOX forms more stable interactions with the micelle than with water. This indicates that DOX encapsulation in PEG-PNIPAM micelles is driven by the drug's tendency to escape water and its stronger interactions with the micelle. This is consistent with the simulation results, where, at equilibrium, most drug molecules are loaded into the micelle, except for PEG20NIP25, where Micelle-W interactions remain more favorable, making the aqueous solution more conducive for micelle dissolution. Previous studies have shown that interactions between drugs and surfactants contribute to micellar solubilization and stabilize incorporated solutes [33]. Our findings provide further evidence for this mechanism. DOX interacts more strongly with micelles in PEG10NIP40 than in other systems, as indicated by higher energy values. The order of interaction energy for DOX-containing micelles (DOX-Micelle) is: NIP25 < PEG10NIP20 < PEG10NIP25 < PEG20NIP25 < PEG10NIP30 < PEG10NIP40. Systems with no PEG (NIP25) and the smallest micelles (PEG10NIP20) exhibit the lowest interaction energies, which increase with PEG content and chain length. Energy landscapes were further analyzed to assess the impact of polymer chain length on the drug stability during encapsulation (Fig. S.8). The Micelle-Micelle energy analysis shows a clear downward trend, indicating micelle formation and stability. The interaction energies between micelles follow the order: PEG10NIP20 > NIP25 > PEG10NIP25 > PEG20NIP25 > PEG10NIP30 > PEG10NIP40. The relatively small energy values for PEG10NIP40 suggest the strongest adhesion between polymer chains, while PEG10NIP20 exhibits the highest energy values, indicating the weakest interactions. A similar conclusion was drawn in a previous study by Shu et al [92] for PNIPAM-Curcumin systems.

### 3.3.7. Molecular Interaction Patterns in Drug-Loaded Micelles

The RDF plots provide valuable insight into the preferred binding sites of the adsorbed molecules and their host-guest interactions in terms of distance and density. A detailed study of PEG-PNIPAM micelles and DOX interactions in aqueous media sheds light on how they self-organize, aggregate, and encapsulate drugs. We first examined the hydration state of drug-loaded PEG-PNIPAM and DOX in aqueous environments (Fig. 19). Compared to blank micelles, RDF peak positions for drug-loaded micelles remain largely unchanged, with only slight differences in peak intensity. The drug-loaded NIP25 micelle shifts from fully hydrophobic to mildly hydrophilic but remains the least hydrophilic system. After DOX addition, PEG20NIP25, PEG10NIP40, and PEG10NIP25 show a slight decrease in peak intensity due to drug presence on the surface, which reduces water accessibility. However, PEG20NIP25 and PEG10NIP40 still exhibit the highest intensities. The dynamic structure of PEG10NIP20 micelles and the movement of drug molecules toward the PEG10NIP30 surface suggest that drug encapsulation does not significantly affect their peak intensity. Water molecules form hydrogen bonds through hydrophilic groups, maintaining a certain amount of water around the micelles.

To support our conclusions, we analyzed drug-solvent interactions, highlighting their role in enhancing the aqueous solubility of hydrophobic drugs. DOX in all systems forms a hydration shell with varying water structuring, consistent with previous studies [33]. DOX is surrounded by more water than copolymer chains due to its surface orientation. Among the systems, encapsulated DOX in PEG10NIP40 and NIP25 micelles show the highest hydration. PEG10NIP40, with its expanded structure and high SASA, allows more water access, while the structural compactness of NIP25 keeps the drug near the micelle surface, interacting with more water molecules. PEG10NIP30, identified as the unstable system based on the gyration radius, shows moderate drug hydration. In contrast, PEG10NIP20, PEG10NIP25, and PEG20NIP25 exhibit the least hydration, indicating DOX distribution toward the inner regions. This internal diffusion causes significant shape changes in PEG10NIP20. In PEG20NIP25, the higher PEG content shields DOX from water, and Rg analysis suggests that DOX stabilizes the micelle structure.

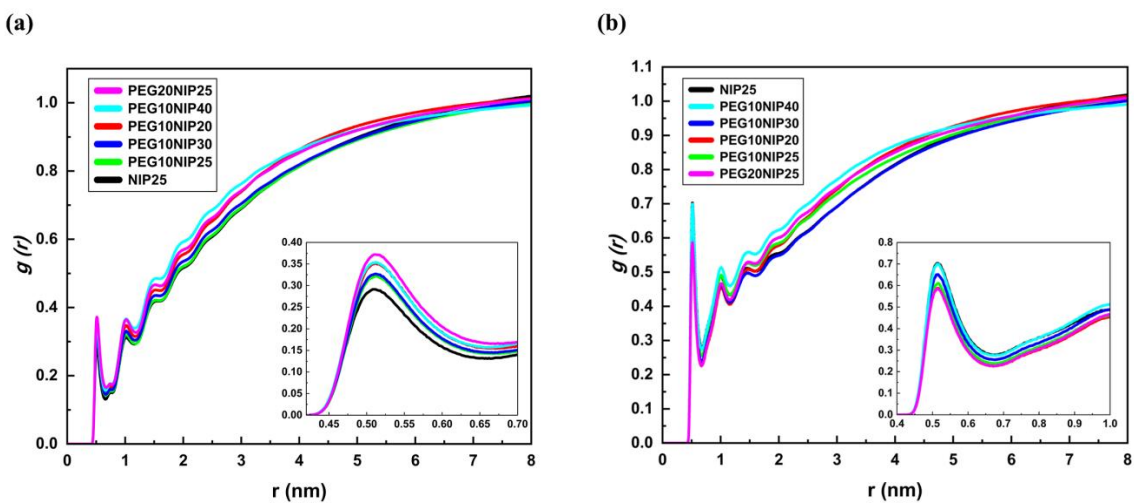

Fig. 19. RDF for water with (a) drug-loaded PEG-PNIPAM and (b) encapsulated Doxorubicin in aqueous environments.

RDF analysis offers insights into drug-micelle interactions by characterizing the distribution of DOX beads around PEG-PNIPAM. The RDF plots provided in Fig. 20 reveal similar molecular orientations across all PEG-PNIPAM-DOX systems, with differences only in peak intensity. The strongest DOX-micelle interactions occur at 0.50–0.55 nm, followed by smaller peaks at 1 and 1.5 nm. Interestingly, these additional peaks are similar for all structures. This suggests that DOX molecules concentrate around PEG-PNIPAM copolymer chains. The steep slope of the RDF plot indicates that the entire DOX molecule inside the box, including those at greater distances, is attracted to the micelle. PEG10NIP20 shows the highest adsorption intensity, suggesting slow DOX diffusion within this model. Systems with higher adsorption intensities (i.e., PEG10NIP30, PEG10NIP40, and PEG20NIP25) exhibit more binding and adsorption stability for DOX molecules. Therefore, RDF data suggest that DOX molecules will have controlled diffusion/transport in these systems. NIP25 micelles show the lowest intensity, indicating that DOX molecules interact most strongly with the PEG-PNIPAM copolymer chains compared to the NIPAM chains. Further analysis of drug dynamics using MSD and diffusion coefficients is provided in Section 3.3.8.

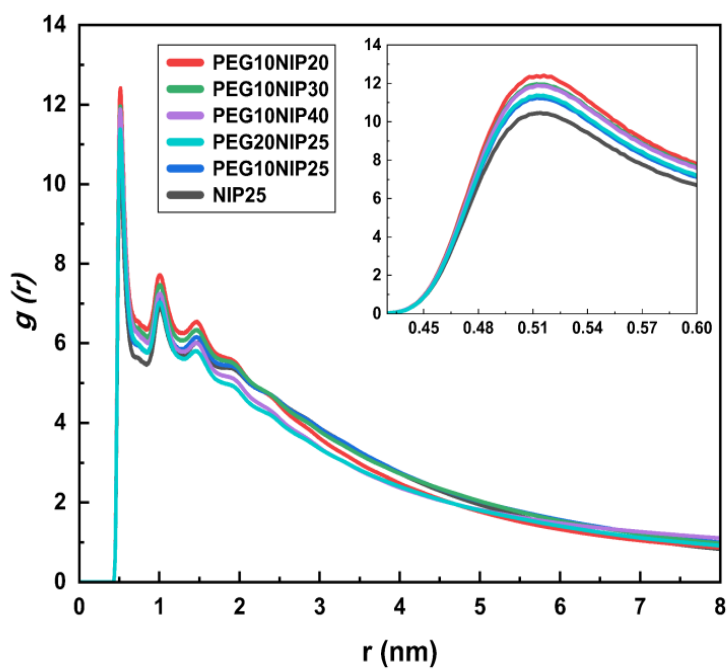

Fig. 20. RDF of drug-loaded PEG-PNIPAM and encapsulated DOX in aqueous environments.

Table 6. Diffusion coefficient of DOX in PEG-PNIPAM micelles

| Samples | Diffusion Coefficient (cm$^2$.s$^{-1}$) |
|---|---|
| NIP25 | $9.01\times10^{-9}$ |
| PEG10NIP20 | $3.53\times10^{-8}$ |
| PEG10NIP25 | $8.32\times10^{-8}$ |
| PEG10NIP30 | $9.27\times10^{-8}$ |
| PEG10NIP40 | $3.28\times10^{-8}$ |
| PEG20NIP25 | $1.95\times10^{-8}$ |

### 3.3.8. Molecular Mobility and Drug Diffusion Dynamics

The segmental block of copolymers plays a crucial role in drug loading, but its impact on drug release and diffusion is less well understood. This section investigates how both hydrophilic and hydrophobic blocks influence the movement, diffusion, and release of drugs from micelles. MSD values offer a visual representation of drug migration and release, while also accounting for any drugs that remain trapped. However, MSD analysis offers further insight into how the polymer blocks behave during drug release. To calculate this, the diffusion coefficient (DC) of DOX in three-dimensional PEG-PNIPAM systems was determined using Einstein's relation in Eq. 2, with the results presented in Table 6 [33,97]:

$$D = \frac{1}{6}\lim_{t\to\infty}\frac{\langle(r(t)-r(0))^2\rangle}{t} \qquad (2)$$

In this equation, the diffusion coefficient ($D$) is derived from the time evolution of the MSD, which represents the average square displacement of a particle over time. Due to the lack of experimental data for the systems studied, the DCs obtained for DOX in PEG-PNIPAM systems were compared with experimental and numerical data for similar DDSs (based on PNIPAM and/or DOX), confirming a consistent order of magnitude ($10^{-8}$ to $10^{-9}$) [98–100]. These results verify that the DCs calculated for DOX in PEG-PNIPAM systems are both acceptable and comparable. Lower DC values suggest stronger drug-micelle interactions. Three main factors influence DC: (1) nanocarrier-drug interaction strength, (2) nanocarrier surface chemistry, and (3) drug molecular weight [101].

The calculated MSD of DOX, as shown in Figure 21, provides insights into viscosity, elasticity, and structural evolution during DOX adsorption onto PEG-PNIPAM micelles. As micelles form and DOX solubilizes, drug diffusivity decreases. Initially, DOX diffuses as individual beads and then diffuses together with the block copolymer micelle after encapsulation. The nearly linear

MSD plots indicate constant diffusion of DOX molecules in DDSs. Among the systems, PEG10NIP25 and PEG10NIP30 exhibit the highest MSD for DOX diffusion, following the trend: PEG10NIP40 > PEG10NIP20 > PEG20NIP25 > NIP25. PEG20NIP25 and NIP25 offer the most controlled DOX transport, making them more suitable for sustained drug delivery. According to Fig. 21, diffusion is easier through the outer layer of PEG10NIP30 than its inner layers. Larger PNIPAM: PEG ratios enhance DOX mobility, while longer PEG chains facilitate controlled transport through electrostatic and hydrogen-bond interactions [102]. Increasing PNIPAM block length densifies the hydrophobic core, expands its surface, and allows more drug distribution at the interface. This balance is essential for drug loading.

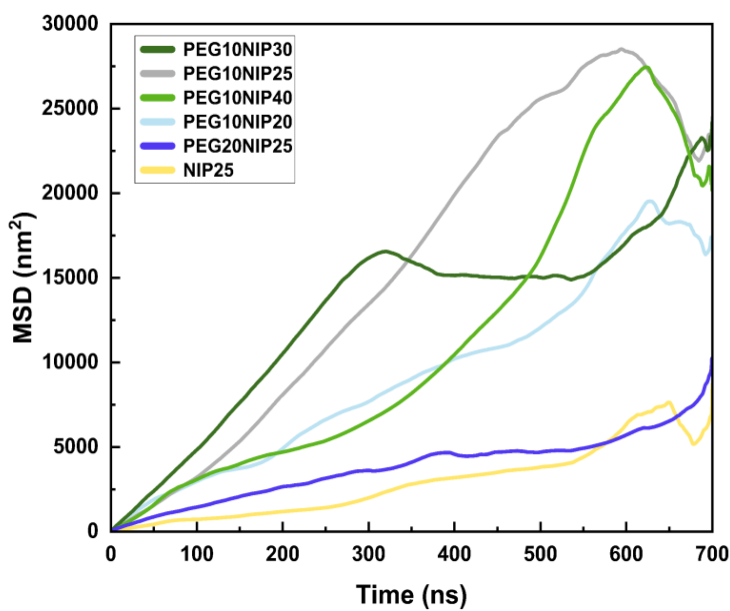

Fig. 21. MSD profile of DOX as a function of time in PEG-PNIPAM micelles.

The diffusion coefficients in PEG10NIP20 reflect strong DOX interactions with the micelle, leading to slower release, while those in PEG10NIP40 and PEG10NIP25 indicate looser structures with more porosity and lower interactions. Two main mechanisms govern drug diffusion in the polymer: the free volume theory, where polymer holes act as transport channels for small molecules, and the wriggling of polymer chains, which aids diffusion by pushing molecules toward areas of lower concentration. Moreover, release time correlates with drug diffusion within micelles [97]. Small diffusion coefficients explain the prolonged release of DOX, so we expect DOX to be slowly released from the PEG20NIP25, NIP25, PEG10NIP40, and PEG10NIP20 micellar systems. The sharp increase in DOX diffusion at the end of the simulation is due to the formation of large pores in the micelle network, transitioning from a uniform structure to cluster/channel morphology. In hydrogels with high water content, DOX molecules reside in these pores, diffusing freely through water-filled spaces [103]. In contrast, the increased PEG block size in PEG20NIP25 restricts DOX movement, causing slower diffusion and larger fluctuations, likely because the flexible blocks limit free movement [104]. Simulation snapshots show DOX molecules adsorbing onto PEG20NIP25 and embedding in chain clusters, confirmed by analyzing the MSD of individual DOX molecules. Polymers with longer hydrophilic blocks form more stable micelles due to a thicker hydrophilic shell. As a result, drugs must pass through this thicker shell before release, which slows down the drug release or causes less apparent change, especially when more drugs are distributed in the core.

Comparing PEG10NIP30 with larger chains reveals that the DC of PEG10NIP40 is smaller than that of PEG10NIP30, which can be explained by the larger order of chains in PEG10NIP30 and its more compact structure compared to PEG10NIP40. This could be further supported by interaction energy for the copolymer chains, which showed that increasing polymer size promotes favorable drug-polymer interactions, making the polymer ideal for constructing DDSs. In summary, segmental block length is connected to drug release, with varying affinities between the drug and micelle following similar release trends. The hydrophilic block shell not only ensures micelle stability but also influences drug release efficiency. Therefore, when designing drug-loaded micelles, the overall impact of segmental block length, including its role in stability and drug release, should be carefully considered.

## 4. Conclusion

This study offers a detailed molecular-level investigation into the self-assembly and drug encapsulation behavior of PEG-PNIPAM micelles for Doxorubicin (DOX) delivery. Using coarse-grained molecular dynamics (CG-MD) simulations, we analyzed the influence of polymer chain length on micelle stability, morphology, hydration, and drug release properties. The results highlight the critical role of polymer composition, particularly the hydrophilic/hydrophobic balance, in determining micelle formation, size, and stability, which in turn govern their performance in drug delivery. Key findings include:

- Systems with shorter PNIPAM segments (PEG to PNIPAM ratios of 0:25, 10:25, and 10:30) formed compact micelles, while longer PNIPAM segments led to elongated structures with reduced packing density.

- PEG incorporation enhanced hydration and steric stabilization, reducing aggregation and improving circulation potential. Increasing PEG length (e.g., PEG20NIP25) led to more hydrated, expanded, and stable micelles, enhancing drug diffusion.

- Increasing PNIPAM content resulted in larger, expanded micelles with stronger hydrophobic core formation, improving drug retention. However, PNIPAM-rich micelles showed limited diffusion pathways, potentially restricting drug release.

- Micelle formation was fastest in systems with shorter PEG chains and dominant PNIPAM segments (e.g., NIP25, PEG10NIP20, PEG10NIP25), suggesting that hydrophobic interactions drive rapid aggregation.

- Thermodynamic and structural analyses confirmed the spontaneous nature of drug encapsulation, driven by hydrophobic interactions and PEG-mediated stabilization. DOX preferentially localized in the PEG regions, benefiting from the hydrophobic PNIPAM core while remaining accessible to the hydrophilic PEG shell, thereby improving solubility and preventing aggregation.

- PEG-rich micelles facilitated sustained drug release due to enhanced hydration, while lower PEG content resulted in higher DOX diffusion coefficients, suggesting weaker retention and faster drug release, which is advantageous for rapid therapeutic action.

- PNIPAM-rich micelles ensured stronger drug retention by providing a more hydrophobic environment, leading to enhanced micelle-drug interactions and delayed DOX release due to hydrophobic trapping.

- Among the systems tested, PEG20NIP25 exhibited optimal stability, making it ideal for sustained drug release, while PEG10NIP30 enabled rapid diffusion, supporting fast-acting chemotherapy applications. PEG10NIP40, with its large core and moderate polymer-water interactions, demonstrated high drug encapsulation efficiency and controlled diffusion, making it a suitable candidate for extended-release formulations.

A key advantage of PEG-PNIPAM micelles is their thermo-responsive nature, which enables temperature-dependent drug release. By fine-tuning the polymer block lengths and the hydrophilic/hydrophobic ratio, these micelles can be optimized for specific drug delivery applications, ensuring both stability and targeted release. This study provides a comprehensive framework for designing next-generation polymeric drug carriers with tunable release kinetics. Future work should extend simulation times to better investigate release properties. Additionally, further research is needed to understand the effects of temperatures above and below PNIPAM's LCST on micellar self-assembly. Subsequent studies could examine how self-assembly, DOX encapsulation and release, and variations in polymer chain length respond to temperature changes. A promising area for future research is exploring the impact of higher drug payloads on PEG–PNIPAM micelles. Investigating how increased drug loading influences micelle stability, release dynamics, and biodistribution could yield critical insights for optimizing therapeutic efficacy while minimizing side effects. Finally, *in vitro* and *in vivo* validation studies are essential to refine these findings and facilitate the clinical translation of PEG–PNIPAM micelles as precision nanocarriers for cancer therapy.